%% file: main.tex
\documentclass[letterpaper,twocolumn,10pt]{article}
\usepackage{titlesec}
\usepackage{siunitx}
\usepackage{url}
\usepackage{usenix}
\usepackage{epsfig}
\usepackage{booktabs}
\usepackage{xspace}
\usepackage{multirow}
\usepackage{listings}
\usepackage{amsmath}
\usepackage{subcaption}
\usepackage{caption}
\usepackage{filecontents}
\usepackage{xurl}
\usepackage{ulem}
\usepackage{wrapfig}
\usepackage{pifont}
\usepackage{comment}
\usepackage{xcolor}
\usepackage[misc]{ifsym}
\usepackage[linesnumbered, ruled]{algorithm2e}
\usepackage[affil-it]{authblk}
\usepackage[justification=centering]{caption}
\usepackage{balance}

\input{tex/setting.tex}
\begin{document}
\date{}

\title{\Large \bf \sysname: Serving Disaggregated Large Language Model at Scale}
\input{tex/author}
\maketitle

\subsection*{Abstract}
Serving disaggregated large language models (LLMs) over tens of thousands of xPU devices (GPUs or NPUs) 
with reliable performance faces multiple challenges.
1) Ignoring the diversity (various prefixes and tidal requests),
treating all the prompts in a mixed pool is inadequate.
To facilitate the similarity per scenario and minimize the inner mismatch on P/D (prefill and decoding) processing,
fine-grained organization is required,
dynamically adjusting P/D ratios for better performance.
2) Due to inaccurate estimation on workload (queue status or maintained connections),
the global scheduler easily incurs unnecessary timeouts in prefill.
3) Block-fixed device-to-device (D2D) KVCache transfer over cluster-level RDMA 
(remote direct memory access) fails to achieve desired D2D utilization as expected.
To overcome previous problems,
this paper proposes an end-to-end system \sysname,
complying with the paradigm of MLOps (machine learning operations),
which models end-to-end (E2E) P/D performance and enables:
1) fine-grained P/D organization,
mapping the service with RoCE (RDMA over converged ethernet) as needed,
to facilitate similar processing and dynamic adjustments on P/D ratios;
2) on-demand forwarding upon rejections for idle prefill,
decoupling the scheduler from regular inaccurate reports and local queues, to avoid timeouts in prefill;
and 3) efficient KVCache transfer via optimized D2D access.
\sysname is implemented upon Ascend and MindSpore,
has been deployed over tens of thousands of NPUs for more than eight months in commercial use,
and further achieves 60\%, 42\% and 46\% improvements 
on E2E throughput, time-to-first-token (TTFT) SLO (service level objective) and D2D transfer time.
As the E2E system with optimizations, \sysname achieves 6.7x increase on throughput,
compared with aggregated LLMs.

\input{tex/introduction}
\input{tex/background}
\input{tex/design}
\input{tex/evaluation}
\input{tex/related}
\input{tex/extension}
\input{tex/conclusion}

\balance
\bibliography{main}
\bibliographystyle{unsrt}

\end{document}

%% file: tex/setting.tex
\makeatletter
\renewcommand\AB@affilsepx{\quad \protect\Affilfont}
\def\thanks#1{\protected@xdef\@thanks{\@thanks
        \protect\footnotetext{#1}}}
\makeatother

\definecolor{codegreen}{rgb}{0,0.6,0}
\definecolor{codegray}{rgb}{0.5,0.5,0.5}
\definecolor{codepurple}{rgb}{0.58,0,0.82}
\definecolor{backcolour}{rgb}{0.95,0.95,0.92}
\lstdefinestyle{mystyle}{
    backgroundcolor=\color{backcolour},   
    commentstyle=\color{codegreen},
    keywordstyle=\color{magenta},
    numberstyle=\tiny\color{codegray},
    stringstyle=\color{codepurple},
    basicstyle=\ttfamily\footnotesize,
    breakatwhitespace=false,         
    breaklines=true,                 
    captionpos=b,                    
    keepspaces=true,                 
    numbers=left,                    
    numbersep=5pt,                  
    showspaces=false,                
    showstringspaces=false,
    showtabs=false,                  
    tabsize=2
}
\SetKwFunction{FunctionName}{FunctionName}
\SetKwProg{Fn}{Function}{:}{}
\SetKwRepeat{Do}{do}{while}
\newcommand{\sysname}{P/D-Serve\xspace}
\newcommand{\li}{\uline{\hspace{0.5em}}}

\newcommand{\authorspace}{\hspace{-1ex + 1.5pt}}

\usepackage{tikz}
\usetikzlibrary{fadings}
\usetikzlibrary{shapes}
\usetikzlibrary{shadows.blur}
\newcommand*\circled[1]{\tikz[baseline=(char.base)]
{\node[shape=circle, color=black, fill=black,  text=white, draw,inner sep=0.4pt] (char) {\bfseries #1};}}

%% file: tex/author.tex
\author[* \authorspace\thanks{$^*$Contributed Equally, Corresponding to jinyibo1@huawei.com.}]{Yibo Jin}
\author[* \authorspace]{Tao Wang}
\author[ \authorspace]{Huimin Lin}
\author[ \authorspace]{Mingyang Song}
\author[ \authorspace]{Peiyang Li}
\author[ \authorspace]{Yipeng Ma}
\author[ \authorspace]{Yicheng Shan}
\author[ \authorspace]{Zhengfan Yuan}
\author[ \authorspace]{Cailong Li}
\author[ \authorspace]{Yajing Sun}
\author[ \authorspace]{Tiandeng Wu}
\author[ \authorspace]{Xing Chu}
\author[ \authorspace]{Ruizhi Huan}
\author[ \authorspace]{Li Ma}
\author[ \authorspace]{Xiao You}
\author[ \authorspace]{Wenting Zhou}
\author[ \authorspace]{Yunpeng Ye}
\author[ \authorspace]{Wen Liu}
\author[ \authorspace]{Xiangkun Xu}
\author[ \authorspace]{Yongsheng Zhang}
\author[ \authorspace]{Tiantian Dong}
\author[ \authorspace]{Jiawei Zhu}
\author[ \authorspace]{Zhe Wang}
\author[ \authorspace]{Xijian Ju}
\author[ \authorspace]{Jianxun Song}
\author[ \authorspace]{Haoliang Cheng}
\author[ \authorspace]{Xiaojing Li}
\author[ \authorspace]{Jiandong Ding}
\author[ \authorspace]{Hefei Guo}
\author[ \authorspace]{Zhengyong Zhang}

\affil[ ]{Huawei Technologies Co., Ltd.}

%% file: tex/introduction.tex
\section{Introduction}
\label{sec:introduction}

Large language models (LLMs) have been widely adopted
~\cite{chatgpt4, team2023gemini, llama3, ren2023pangusigma, kimi} 
for various generative applications.
In order to shorten time-to-first-token (TTFT) for quick response
(e.g., first token returned to users immediately, followed by further streaming results) 
and enhance the throughput of decoding follow-up tokens at the same time, 
disaggregated paradigm has become a new trend
~\cite{patel2023splitwise, cunchen2024tetriinfer, gao2024attentionstore, qin2024mooncake, zhong2024distserve}, 
where the prefill phase and decoding phase are deployed in different instances, 
with disparate settings on batch sizes.
Then, the transfer of intermediate data, generated during inference (i.e., KVCache), is necessary.

Although previous works have already implemented disaggregated LLMs upon vLLM~\cite{llm_vllm_2023},
using pool management for prefill and decoding instances,
and further instance migration and role switch for various workloads~\cite{patel2023splitwise, cunchen2024tetriinfer},
serving disaggregated LLMs at scale, 
over tens of thousands of xPU devices (NPUs or GPUs),
faces multiple challenges as follows:

First and foremost, ignoring the diversity,
treating all the prompts in a mixed pool is inadequate.
Given the P/D (prefill and decoding) instances, 
end-to-end (E2E) throughput depends on its bottleneck
(e.g., longer prompts slow down the prefill while more tokens generated increase the occupation in the decoding).
Faster inference is preferred,
but minimizing the P/D mismatch is more urgent (maximum capability of serving).
As mentioned in~\cite{gim2023prompt, ye2024chunkattention},
the inference can be accelerated by reusing attention states across prompts (i.e. prefix-aware KVCache).
However, the weights and all KVCaches share the high bandwidth memory (HBM).
With the growth of model size and prompt length,
KVCache dramatically increases~\cite{kwon2023pageattention},
implying that individual instance in the pool fails to cover all the prefixes for all scenarios.
Except for using host memory as substitute (incurs load and flush),
assigning the prompts with common prefixes
to a few prefills enhances the hit rate.
Although prefill is accelerated, without further adjustments in decoding
(e.g., larger batch size upon shared prefixes and cost-effective P/D ratio),
the throughput is still limited.
Actually, processing homologous prompts together is profit to apply more optimizations,
but fine-grained orchestration should be considered
(e.g., changes on P/D ratios due to updated prefixes via prompt engineering
~\cite{jason2022cot, white2023promptpattern}).

Furthermore, 
the global scheduler fails to balance the workload among prefill due to inaccurate estimation.
As shown in previous works~\cite{patel2023splitwise, cunchen2024tetriinfer},
each prefill instance regularly communicates to the scheduler
(e.g., reporting the queue every 100ms).
Such queue status based on either pending tokens or memory capacity is actually insufficient,
since the inference in prefill also depends on the length of cached prefix and batch size.
Given TTFT SLO (service-level objective) per scenario, 
hitting a longer prefix implies a larger batch size can be applied.
With the changes on prefix and batch size, actual TTFT has great varying range.
Although the gateway records the server-sent events (SSE) connections for streaming responses,
the workload hinted by the number of connections also fails to indicate idle prefill,
since it is maintained during entire LLM lifecycle (also including decoding).
Current scheduler fails to capture these factors, leading to sub-optimum.
Instead of directly forwarding requests to the queue of each prefill (easily incurs timeouts),
a more effective way is: 
the requests have the chance to be assigned to other idle prefill.

\setlength{\parskip}{0.9pt}
In addition,
block-fixed device-to-device (D2D) KVCache transfer 
over cluster-level RDMA (remote direct memory access) fails to achieve desired performance.
PageAttention~\cite{kwon2023pageattention} is widely used
for efficient management of xPU memory and thus enhance the throughput in decoding phase.
Although the optimizations like KVCache transfer per layer are adopted~\cite{patel2023splitwise},
when both sender and receiver are equipped with PageAttention,
KVCache transfers between P/D over RDMA are naturally implemented using discrete blocks (e.g., transfer one by one).
Such transfer incurs unnecessary software overhead
and fails to fully utilize D2D bandwidth.
Note that transfer per block involves the confirmation between the sender and the receiver,
where the controls actually waste the bandwidth.
Thus, it is preferred to transfer all the blocks as a whole (in bytes),
involve less controls,
and recover the bytes to blocks in decoding.
It is also challenging to ensure contiguous space at sender
as well as to make the tradeoff between transparency and flexibility for restoring discrete blocks.
Moreover, multiple hops may incur the conflicts and unstable transfer time,
which requires the platform to fully utilize the path diversity.

Existing research falls insufficient for addressing the aforementioned challenges.
Some works had studied accelerating LLMs via various caching, management and kernel optimizations
~\cite{llm_vllm_2023, kwon2023pageattention, agrawal2023sarathi,
nvidia_fastertransformer_2023, dao2023flashattention2, gim2023prompt, ye2024chunkattention, agrawal2024taming}.
Others focused on the batch scheduling (e.g., continuous batching) within and over instances
~\cite{aminabadi2022deepspeed, yu2022orca, wu2023fast, li2023alpaserve, sheng2024fairness}.
And the rest investigated serving the disaggregated LLMs for better performance
~\cite{patel2023splitwise, cunchen2024tetriinfer, gao2024attentionstore, qin2024mooncake}.
However, few of them has considered serving E2E disaggregated LLMs at scale,
over tens of thousands of xPU devices (NPUs or GPUs),
which involves new challenges on both processing and scaling.

In this paper,
we propose an end-to-end system \sysname for serving disaggregated LLMs at scale. 
Firstly, \sysname models the E2E P/D performance.
To facilitate similar processing on homologous prompts and on-demand adjustments,
\sysname uses dynamic mappings of 
the services (or further scenarios) and the RoCE (RDMA over converged ethernet) as needed,
where RoCE supports D2D KVCache transfer.
Via dynamic RoCE mapping, 
fine-grained organization is then enabled (for specific scenario in a service),
including rolling upgrade, scaling upon groups,
and the adjustment on P/D ratios catering to content changes and traffic changes.
Secondly, based on such fine-grained organization for P/D instances,
\sysname uses a customized monitor for auto health check of xPU device.
Meanwhile, the auto recovery is performed efficiently,
which only substitutes the fault one with minimum cost and does no harm to running service.
Thirdly, to pursue balanced workload among prefills,
\sysname uses on-demand forwarding upon the rejections for idle prefill,
in which the requests have the chance to be assigned to further idle one by gateway retries,
decoupling the scheduler from regular inaccurate reports and the local queues.
Meanwhile, the gateway also enables batch forwarding, 
catering to various batch settings among prefills.
Finally, to fully utilize D2D bandwidth,
\sysname manages the KVCache to be transferred via a contiguous buffer,
and recovers the bytes to desired discrete blocks via RecvScatter.
Meanwhile, stable transfer and RoCE construction are ensured for large scale xPUs.
\sysname is implemented upon Ascend and MindSpore,
has been deployed over tens of thousands of NPUs for more than eight months in commercial use,
and further achieves 60\%, 42\% and 46\% improvements 
on E2E throughput, TTFT SLO and D2D transfer time.
Compared with serving aggregated LLMs,
\sysname achieves 6.7x increase on throughput at scale.

%% file: tex/background.tex
\section{Background and Motivation}
\label{sec:background}

This section first introduces the fundamentals of autoregressive LLM,
and then explicitly show the performance degradation in production,
even using the disaggregated paradigm.
Upon the degradation,
this section discusses the opportunities and challenges
for serving disaggregated LLM at scale.

\subsection{Autoregressive LLM}
\textbf{LLM}:
The architecture of transformer~\cite{vaswani2023attentionneed} 
stands on two pillars: the encoder and the decoder,
in which the self-attention mechanism is the core.
The encoder and decoder can be used in combination or separately.
A typical example of encoder-only model is Bert~\cite{devlin2018bert}
while an example of decoder-only model is Llama~\cite{llama3}.
The input of LLM inference is a prompt 
(a piece of text, converted to tokens in advance via a tokenizer)
while the output is a sequence of tokens.
The tokens are generated in an autoregressive manner.
That is, the LLM is trained on a vast collection of texts,
with the goal of predicting the next token based on the previous ones.
The prefill refers to the inference on predicting the first token
while the decoding refers to the inference on predicting the follow-up tokens.
The metric of prefill phase is time-to-first-token (TTFT),
and the metrics of decoding phase are time-per-output-token (TPOT).
For entire LLM, the metrics are the latency to generate all tokens (E2E time or latency),
and the throughput.

\textbf{KVCache}:
The attention block of transformer (e.g., multi-head self-attention) 
essentially contains scaled dot-product attention,
where the previous outputs are represented into a query (\textbf{Q}) tensor
and the next output is produced by mapping this query and 
the set of keys (\textbf{K}) and values (\textbf{V}).
Since the previous tokens are appended one after another to form \textbf{Q},
to reuse the intermediate data (only incurs the computation for newly appended one),
already computed key-value pairs (\textbf{K} and \textbf{V}) are cached (i.e., KVCache).
The size of current KVCache per layer is calculated by
2 (bytes for fp16) * batch size * hidden size * 2 (\textbf{K} and \textbf{V} tensors) * query length.
Note that the query length is the sum of tokens in prompt and the tokens already generated.
With the growth of LLM, actually the hidden size and the number of layers,
KVCache increases dramatically.
As shown in~\cite{dettmers2022gpt3},
GPT-3 (175B) generates a 4.5MB KVCache per token.
Then, for prefill inference with the prompt of 1k tokens, KVCache is 4.5GB.
Some techniques 
like quantization~\cite{dettmers2022gpt3,dettmers2022spqr,isik2023gptzip,xiao2023smoothquant} 
and grouped attention~\cite{ainslie2023gqa,tang2024survey}
are used for small KVCache.
With the growth of the model and the context length, 
the increase of KVCache is inevitable.

\textbf{Disaggregated LLM}:
In order to pursue lower TTFT and achieve a higher throughput for follow-up tokens,
disaggregated LLM is proposed,
where prefill phase and decoding phase (P/D) are deployed separately,
decoupling the constraint on the batch size.
That is, the ratio of P/D instances can be configured in advance for various workloads.
For those prompts generating few tokens,
more prefill instances are preferred,
with the inference executed under small batch size and using the pipeline one batch after another.
For those prompts generating a large volume of tokens,
more decoding instances are preferred (or with larger batch size).
And for each prompt, already executed in prefill,
its KVCache has to be transferred to a decoding instance for further inference.
Since the KVCache is stored among multiple xPU devices 
(split by the strategy of model parallelism, e.g., tensor parallelism),
D2D KVCache transfer in order is involved,
where the data stored in the 0-th device of the sender is transferred to
the 0-th device of the receiver correspondingly (multiple transfers simultaneously).

\textbf{Infrastructure}:
Upon the physical machines, 
the containers are used as the minimum resource unit for service scaling,
where each container is assigned multiple xPU (e.g., volcano~\cite{volcano} for NPU) devices for inference 
(prefill phase or decoding phase).
Each xPU device has limited HBM (e.g., tens of GB) and is directly connected to top-of-rack (ToR) switch via RDMA.
ToR switches further connect to spine one.
The maximum RoCE (RDMA over converged ethernet) IPs are limited in a region, in thousands.
The maximum bandwidth for D2D transfer is about hundreds of Gb per second.

\textbf{E2E Performance}:
Except for the E2E latency measuring the sum of prefill and decoding latency,
given P/D instances, the throughput actually depends on the bottleneck:
\begin{gather*}
	\Phi = \min\{I_{t},\;\;\; n_{p}b_{p}\cdot 1/T_{p},\;\;\; n_{d}b_{d}\cdot 1/T_{d}\} / (n_{p} + n_{d}), \\
    \hspace{-2pt}T_{p} = TTFT_{bs} * r_{pre}, \;\; T_{d} = \xi + TPOT_{bs} * G, \;\; E2E = T_{p} + T_{d},
\end{gather*}
where the numerator of $\Phi$ measures the bottleneck (i.e., RPS, requests per second).
$p$ and $d$ refer to ``prefill'' and ``decoding'', respectively.
$n_{p}$ and $n_{d}$ are the number of prefill instances and decoding instances, respectively.
P/D ratio refers to $n_{p}/n_{d}$.
$\Phi$ refers to the average throughput per instance (also measures the cost).
$I_{t}$ refers to the input traffic.
Limited to the bottleneck, not all requests are treated without breaking timeouts.
For example, waiting is involved if prefill is weak: $I_{t} > n_{p}b_{p}\cdot 1/T_{p}$.
Here, $t$ refers to the time slot, since the input traffic varies over time.
$b_{p}$ and $b_{d}$ are the batch sizes used in prefill and decoding.
$1/T_{p}$ and $1/T_{d}$ measure the processing capability in prefill and decoding 
(i.e., averaged batches treated per second; ``1'' refers to a unit time period).
In prefill, $TTFT_{bs}$ refers to the TTFT under batch size $bs$,
and $r_{pre} \in (0, 1]$ refers to the ratio benefit from prefix-aware KVCaches.
In decoding, $\xi$ measures the time of KVCache transfer.
Since each D2D transfer involves multiple sub-transfers,
$\xi$ refers to the maximum one.
Similarly, $TPOT_{bs}$ refers to the TPOT under batch size $bs$.
$G$ refers to the averaged tokens generated in decoding.

\subsection{Opportunities and Challenges}
Serving disaggregated LLMs at scale, for services and over xPUs,
unavoidably faces performance degradation.
And for each opportunity, related challenge is highlighted.

\subsubsection{Diversity on Prompts Affects E2E Latency}
\begin{figure}[!t]
    \begin{subfigure}[h]{0.22\textwidth}
        \setlength{\abovecaptionskip}{2pt}
        \includegraphics[width=1.62in,height=1.265in]{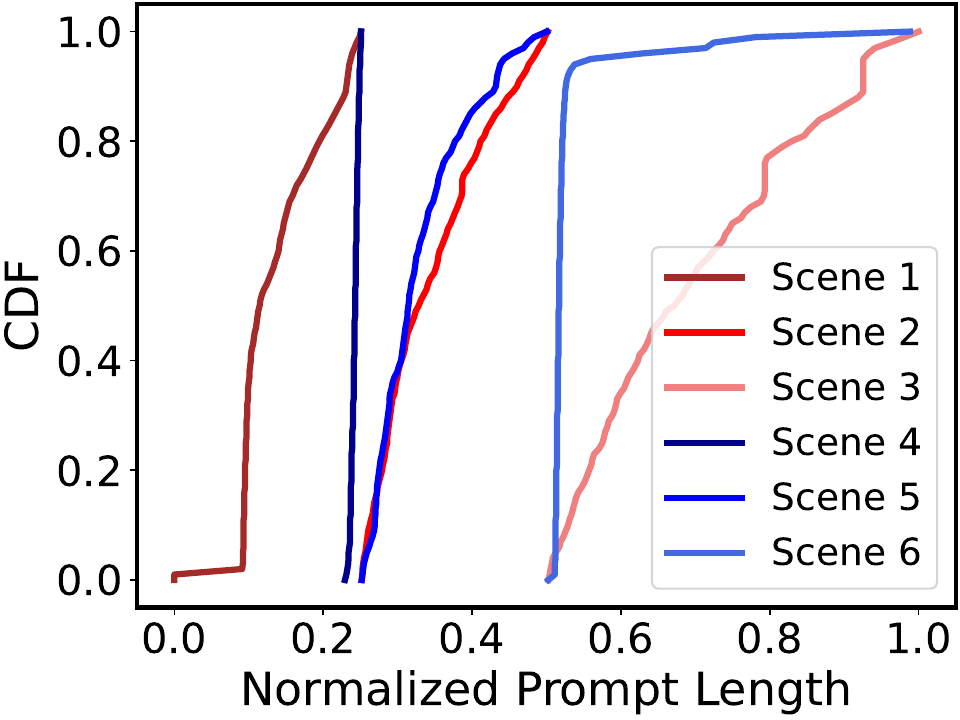}
        \centering
        \caption{Various Prompts \protect\\ in Services (Scenarios)}
        \label{fig:background_diversity_1}
    \end{subfigure}
    \hfill
    \begin{subfigure}[h]{0.22\textwidth}
        \setlength{\abovecaptionskip}{2pt}
        \includegraphics[width=1.62in,height=1.265in]{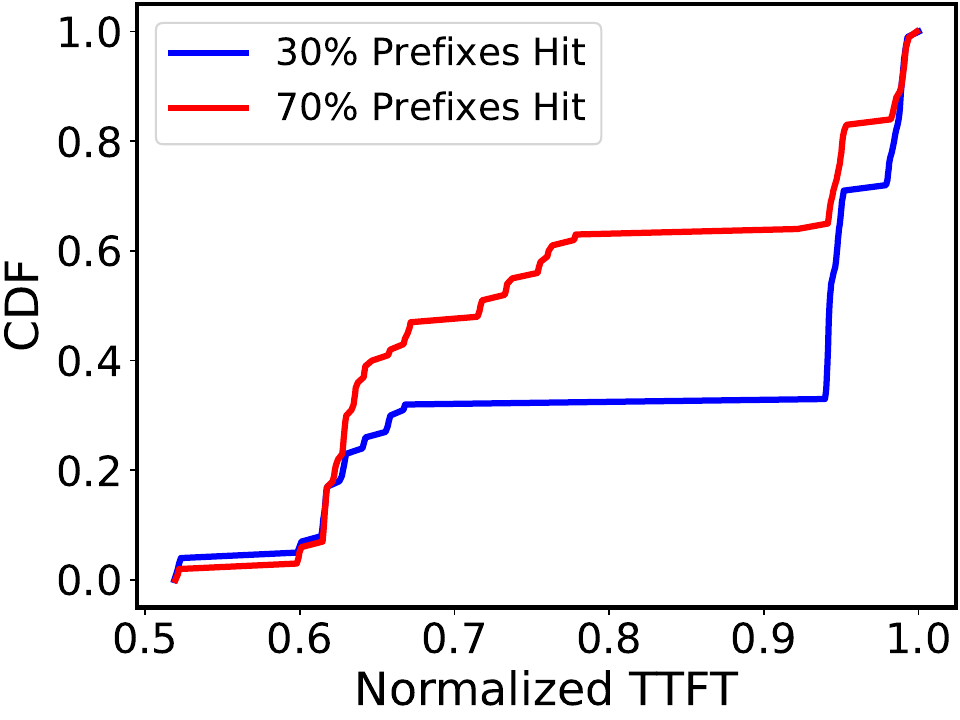}
        \caption{TTFT under Various\protect\\ Prefixes in HBM}
        \label{fig:background_diversity_2}
    \end{subfigure}
    \vspace{-9pt}
    \caption{Performance degradation derives from the diversity.}
    \vspace{-12pt}
\end{figure}

\textbf{Diversity on Prompts}:
The prompt is a piece of text or a set of instructions,
as the input of LLM (converted to tokens in advance via a tokenizer) 
to trigger a specific response or action.
The prompt contains two parts: the setting part and the query part.
The setting part is used to describe the system and the scenario.
For example, Llama 3~\cite{llama3} uses fixed format to indicate the role:
``<|begin\li of\li text|><|start\li header\li id|>\{role\}<|end\li header\li id|>'',
where the role can be system, user or assistant.
In majority scenarios,
the settings are very common to involve the contexts~\cite{yue2023llamarec}.
A typical context is: ``Candidate Pool: (A)...(B)...(C)...'', 
where all candidate options are listed for LLM inference.
Some background facts can also be involved
in either system part and scenario part as auxiliary information.
Actually, the developers from different services or even the scenarios in a service,
individually design the prompts (via prompt engineering).
Although the format is similar,
prompt (or prefix) length is quite different (also the contents), 
as in Fig.~\ref{fig:background_diversity_1}
(there are six scenarios from two services, Scene 1$\sim$6).
And, the traffic also changes over time (i.e., the combination of prompts among scenarios).

\textbf{Limited HBM}:
The memory of xPU device is limited (i.e., the size of HBM is tens of GB).
In order to avoid exchanging the data with host memory,
all the tensors used for inference are preferred to stored in xPU HBM,
including the model weights, the activations, the space left for reserved usage, and the KVCaches.
As shown in~\cite{kwon2023pageattention},
based on Nvidia A100 with 40GB HBM, the space left for KVCache is at least 30\% for 13B LLM.
Although some techniques like quantization and grouped attention are used,
KVCache dramatically increases with the growth of model size and prompt length.
For example,
GPT-3 (175B) generates a 4.5MB KVCache per token.
Using int-8 quantization, the KVCache for 1k prompts is reduced to about 2.3GB.
However, with the growth of prompt length (e.g., 4k), the size dramatically increases to about 9GB.
That is, the KVCache dominates the HBM use.

\textbf{E2E Degradation}:
If all instances are organized in a fixed pool 
(e.g., service-level management: inference at daytime and training at night),
each one of deployed prefill instances has to serve all the possible prompts,
which requires them to cache all the prefix-aware KVCaches~\cite{gim2023prompt,ye2024chunkattention}.
Unfortunately, given limited HBM, caching all the KVCaches per instance is impossible 
(multiple scenarios per service, and tens of prefixes per scenario).
Given HBM, the hit rate of prefix significantly affects TTFT (actually $T_{p}$, with batch processing and cached prefixes),
as in Fig.~\ref{fig:background_diversity_2}.
Although several works~\cite{gao2024attentionstore,qin2024mooncake} 
have already considered the pools for large volume of KVCaches 
with the help of host memory (incurs load and flush),
to fundamentally improve the hit rate of cached prefixes,
increasing valuable KVCaches in HBM is inevitable.

\textit{\textbf{Opportunity}}:
The similarity occurs in P/D behaviors
(e.g., the prompt length or the number of tokens generated 
is more likely to be similar in a scenario),
motivating serving prompts together.
The inference could be further accelerated due to homologous features 
(e.g., similar prompts facilitate input prepare in advance or customized parallelism~\cite{wu2024loong}).
Upon the similarity, 
better performance could be achieved.

\begin{figure}[!t]
    \begin{subfigure}[h]{0.22\textwidth}
        \setlength{\abovecaptionskip}{2pt}
        \includegraphics[width=1.62in,height=1.265in]{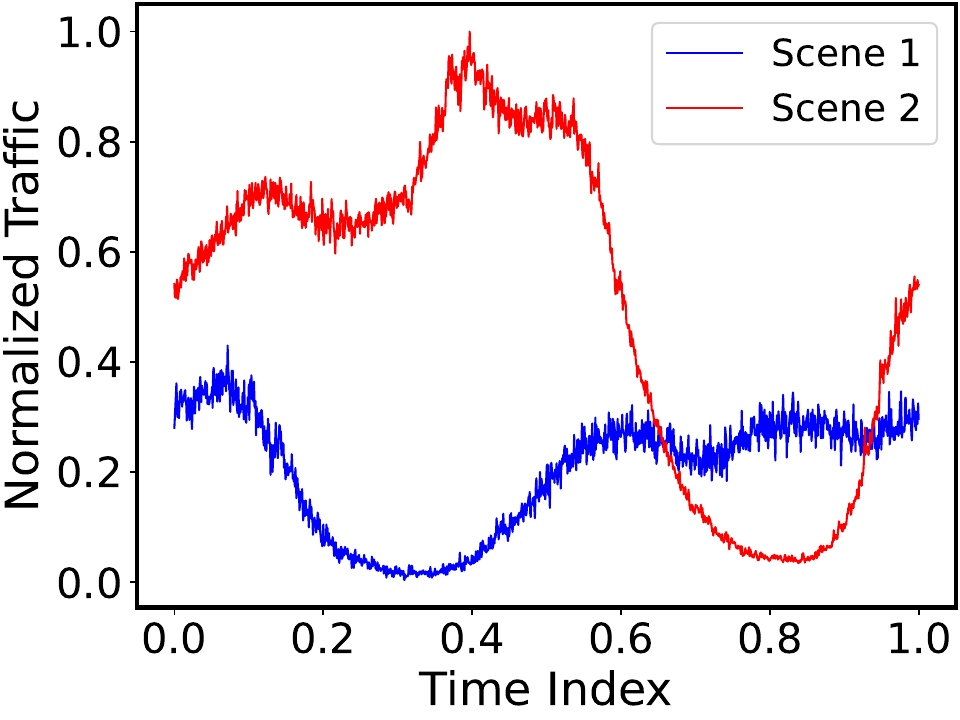}
        \caption{Traffics among Scenarios} 
        \label{fig:background_traffic}
    \end{subfigure}
    \hfill
    \begin{subfigure}[h]{0.22\textwidth}
        \setlength{\abovecaptionskip}{2pt}
        \includegraphics[width=1.62in,height=1.265in]{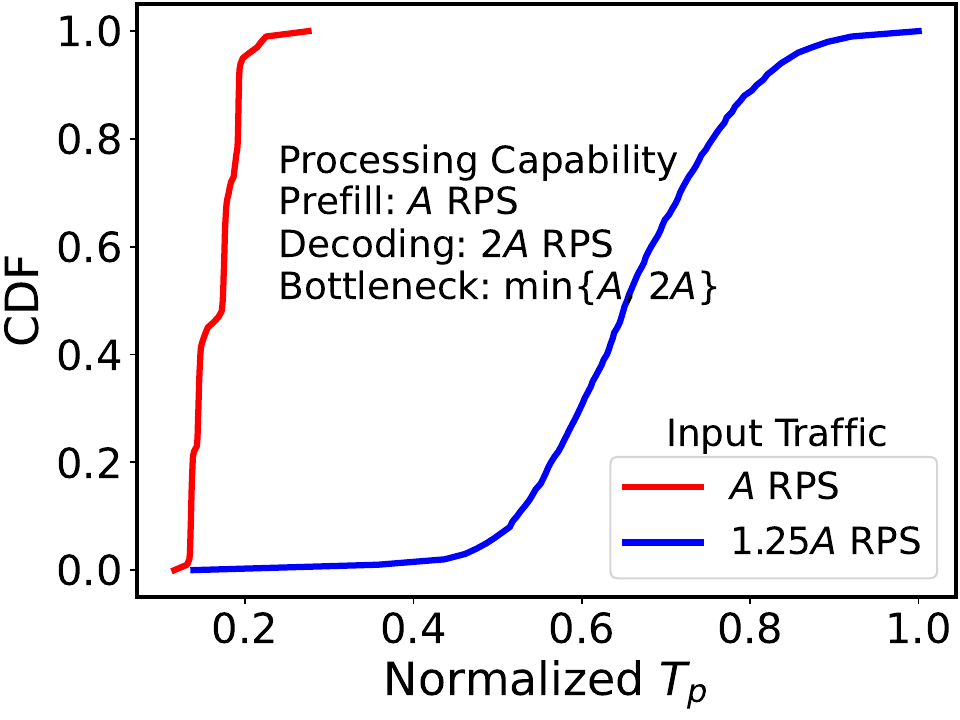}
        \caption{P/D Mismatch}
        \label{fig:background_mismatch}
    \end{subfigure}
    \vspace{-9pt}
    \caption{Changes and Mismatch in Disaggregated LLMs}
    \vspace{-9pt}
\end{figure}

\textit{\textbf{Challenge}}:
The changes on prompts derive from both content and traffic.
Then, corresponding adjustments are needed from two aspects: scaling demands and efficiency demands.
Organizing a subset of prefill instances and related decoding ones,
actually involves complicated management on rolling upgrade of models, prompts, etc., 
and necessary auto scale-out and recovery (minimum cost required).
Furthermore, catering to the content changes (due to prompt engineering) 
and traffic changes (shown in Fig.~\ref{fig:background_traffic}),
orchestrating P/D ratio is a must,
which involves inevitable re-organizing existing P/D groups dynamically.
Essentially, the orchestration has to minimize the mismatch (in Fig.~\ref{fig:background_mismatch})
regarding the processing capability
between prefill and decoding instances.

\subsubsection{Inaccurate Prefill Status Affects TTFT}
\begin{figure}[!t]
    \begin{subfigure}[h]{0.22\textwidth}
        \setlength{\abovecaptionskip}{2pt}
        \includegraphics[width=1.62in,height=1.265in]{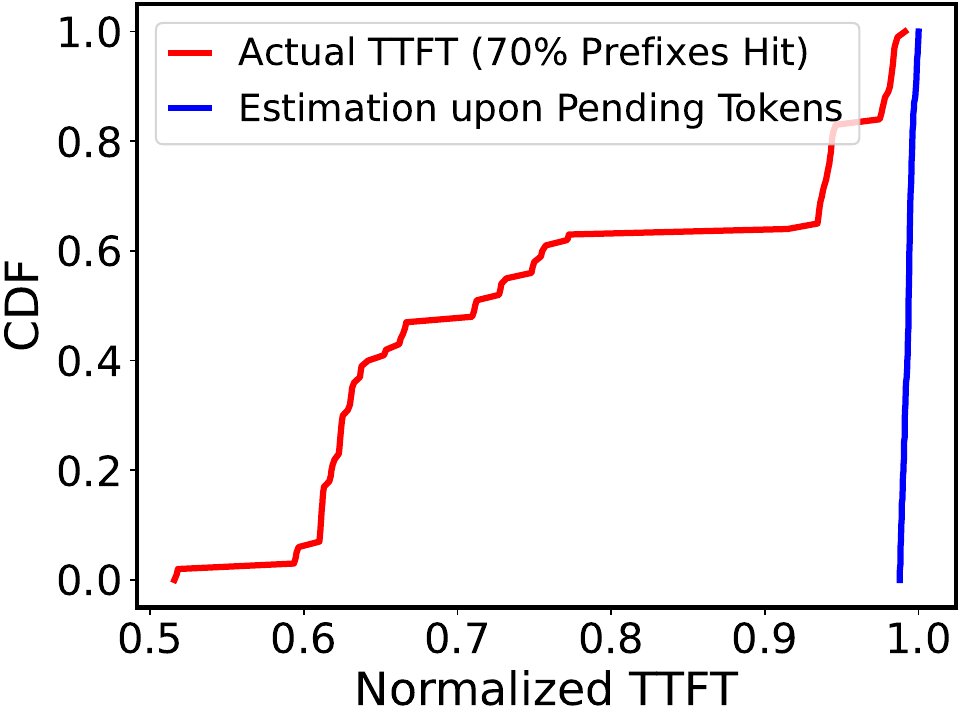}
        \caption{Actual TTFT v.s. \protect\\Token-based Estimation} 
        \label{fig:background_inaccuracy_1}
    \end{subfigure}
    \hfill
    \begin{subfigure}[h]{0.22\textwidth}
        \setlength{\abovecaptionskip}{2pt}
        \includegraphics[width=1.62in,height=1.265in]{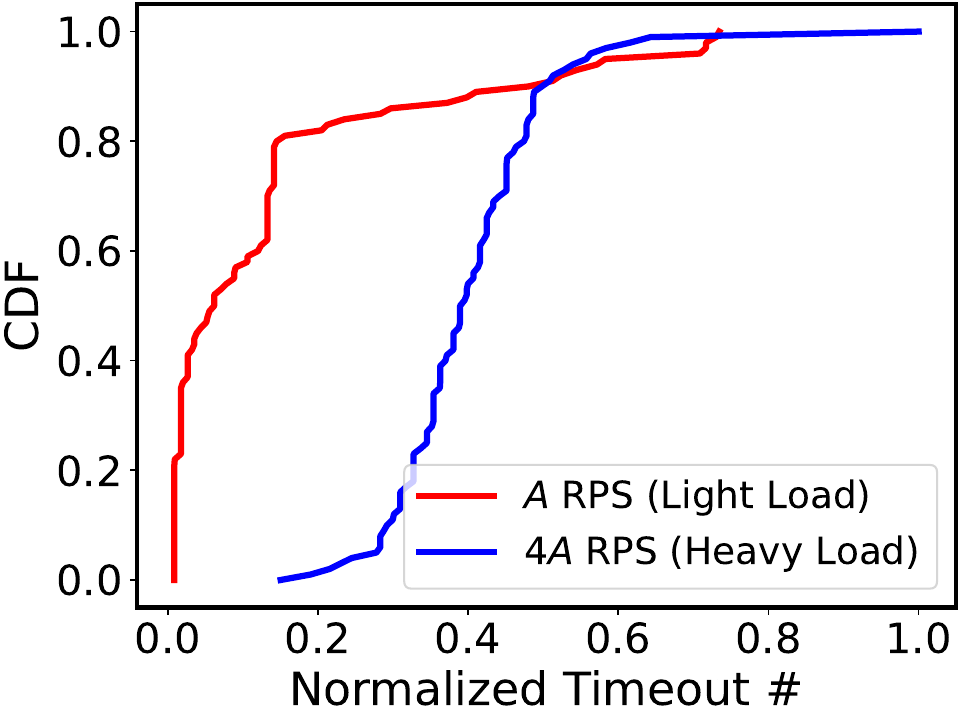}
        \caption{TTFT Timeouts \protect\\due to Prefill Queueing}
        \label{fig:background_inaccuracy_2}
    \end{subfigure}
    \vspace{-7pt}
    \caption{Queue status is insufficient for precise TTFT.}
    \vspace{-11pt}
\end{figure}

\textbf{Inaccurate Queue Status}:
Existing scheduler collects the queue status or available free HBM from the prefill instances,
in which the queue and the free HBM are measured by using pending tokens to be treated.
Unfortunately, those pending tokens are far from precise TTFT prediction.
By using the prefix-aware KVCaches, prefill inference is accelerated.
Thus, more prompts can be treated simultaneously in a single batch,
as long as the TTFT does not exceed a given threshold.
Then, pending tokens along are inadequate.
As in Fig.~\ref{fig:background_inaccuracy_1},
compared with TTFT under 70\% prefixes hit (actually $T_{p}$, with batch processing and cached prefixes),
the estimation (already considered the batch size) upon tokens is inaccurate.
Note that the prompt lengths are similar in this case.
Thus, the gap exists between the estimation (the blue line) and actual TTFT (the red one).
Further, the reports are triggered regularly (e.g., every 100ms).
The period between two consecutive ones also hampers the scheduler from precise decision.

\textbf{Unnecessary Queueing in Prefill}:
Except for inaccurate estimation on TTFT,
queueing in prefill easily incurs the timeouts.
Here is a simple example.
There are two prefill instances (batch size 1 and empty queue).
And four prompts arrive simultaneously.
The length of prompt 1 is 8k and the lengths of the others are 2k.
Both prefill instances contain the prefix-aware KVCaches,
where the prefix for prompt 1 is 0.1k and the prefix for the other prompts is 1k.
Due to simultaneous arrival,
either shortest queue strategy or round robin strategy assigns two prompts to the first prefill instance
and the other two are assigned to the rest.
However, the optimum strategy is to assign three prompts with length 1k to the same instance.
Here, the prompts, waiting in the queue, are affected (i.e., the 1k one queued after 8k).
Although the prompts terminate immediately after the timeouts,
they still affects those queuing ones.
Note that the timeout threshold for 1k is quite different from that of 8k.
As in Fig.~\ref{fig:background_inaccuracy_2},
under heavy workload,
requests are more likely to break timeouts, especially for short prompts.
A more effective way is to assign the prompts only to idle prefill.
After completing two prompts with length 1k,
the third prompt with the same length is then assigned, 
instead of waiting in local queue for the completion of 8k prompt.
That is, waiting in local queue is sub-optimal.

\textbf{TTFT SLO}:
TTFT SLO refers to achieving a higher ratio of the prompts, 
whose TTFT (actually $T_{p}$, with batch processing and cached prefixes) does not exceed related thresholds.
Both prefill and the gateway are configured using early intervention.
As long as current duration exceeds the TTFT SLO (e.g., waiting in the queue or failure),
either prefill or gateway may complete related requests, 
to avoid resource waste and to avoid the bottleneck as early as possible.

\textit{\textbf{Opportunity}}:
For global scheduler, 
inaccurate status upon queues (just pending tokens) or the connections (cover entire LLM lifecycle)
should be fundamentally avoided.
A more effective way is: assigning pending prompts only to idle prefill.

\textit{\textbf{Challenge}}:
Actual TTFT depends on both batch size and prefix-aware KVCaches 
(i.e., $T_{p}$, posterior revealed, and vary among scenarios).
Blindly forwarding new requests to a few idle instances incurs the waiting in a local queue.
And for the traffic surge, those waiting requests easily break the timeout thresholds, 
leading to a lower success rate.
To get rid of the sub-optimum waiting and to avoid the preemptive scheduling,
the queues in prefill are preferred to be removed.
Afterwards, all pending prompts have to be waiting at gateway for further choices.
Unfortunately, lack of accurate feedback from prefill, the scheduler fails to be aware of idle ones in time.
Essentially, global scheduler has to minimize the mismatch 
between the traffic and the processing capability of P/D instances.

\subsubsection{Block-fixed KVCache Affects Transfer}
\textbf{Block-fixed Management}:
Block-fixed memory management has been widely used for xPU HBM 
(e.g., PageAttention~\cite{kwon2023pageattention}),
where the HBM is organized via discrete blocks with fixed size.
Benefit from organizing the HBM in discrete blocks,
more prompts can be involved, fully utilizing the memory,
which facilitates the decoding phase and enhances the throughput.
For aggregated LLM,
vLLM enables PageAttention for entire HBM (both prefill and decoding).
Upon vLLM, previous works have implemented disaggregated LLM.
Both sender and receiver equipped with PageAttention,
KVCache transfers between P/D instances over RDMA 
are naturally implemented using the blocks (e.g., transfer one by one).
However, as shown later, block-fixed transfer is inefficient.

\begin{figure}[!t]
    \begin{subfigure}[h]{0.22\textwidth}
        \setlength{\abovecaptionskip}{2pt}
        \includegraphics[width=1.62in,height=1.265in]{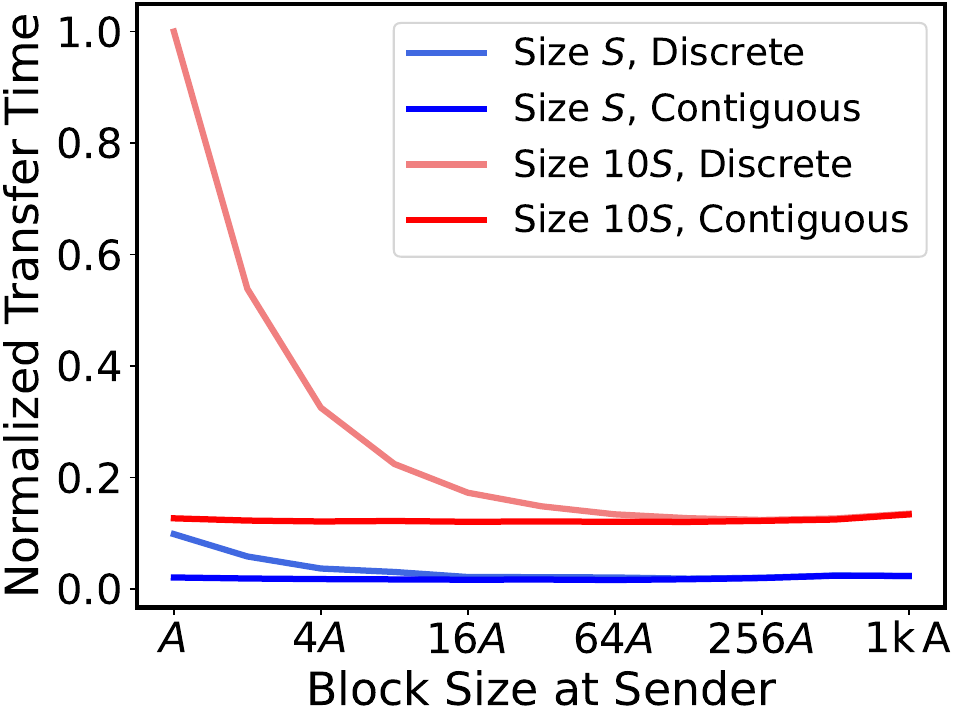}
        \caption{Extra Control Cost} 
        \label{fig:background_block_1}
    \end{subfigure}
    \hfill
    \begin{subfigure}[h]{0.22\textwidth}
        \setlength{\abovecaptionskip}{2pt}
        \includegraphics[width=1.62in,height=1.265in]{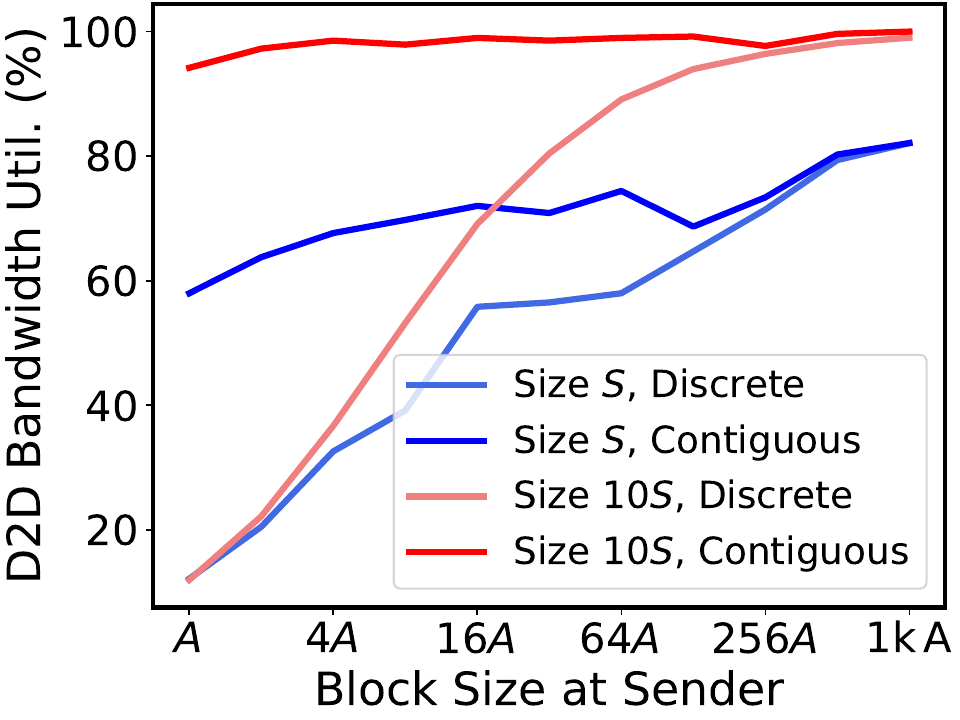}
        \caption{Bandwidth Utilization}
        \label{fig:background_block_2}
    \end{subfigure}
    \vspace{-6pt}
    \caption{Block-fixed transfer fails to fully utilize bandwidth.}
    \vspace{-7pt}
\end{figure}

\textbf{Unnecessary Controls}:
KVCache is the intermediate data, generated during inference,
and is calculated per layer.
Instead of transfer the entire KVCache after the completion of prefill,
transfer per layer is proposed for concurrent computation and transmission.
That is, during the calculation of next layer,
the key-value pairs already generated for previous layer could be transferred.
Naturally, KVCache transfer is implemented using blocks one after another (i.e., discrete).
However, RDMA prefers to transfer all the data as a whole (in bytes, i.e., contiguous),
to avoid frequent confirmations and controls.
As in Fig.~\ref{fig:background_block_1},
with the growth of the data size to be transferred under small blocks,
the extra control cost increases,
which should be avoided (waste D2D bandwidth).
Further results are illustrated in Fig.~\ref{fig:background_block_2},
where the bandwidth utilization is low using discrete blocks, due to frequent controls.

\textbf{Transfer Variance}:
Typically, the entire transfer time is expected to be controlled 
(e.g., a small fraction of TPOT).
Otherwise, there is a noticeable gap between the first token and the follow-up ones.
Furthermore, the transfer should be stable.
Unfortunately, when conflicts occur along with multiple hops,
the transfer time varies dramatically, even leading to hundreds of milliseconds.
Such variance is unacceptable.

\textit{\textbf{Opportunity}}:
Both sender and receiver may manage xPU HBM via PageAttention (i.e., discrete blocks).
To fully utilize D2D bandwidth over RDMA,
transfer in bytes as a whole is preferred, instead of block transfers one after another.

\textit{\textbf{Challenge}}:
To facilitate block-free KVCache transfer, 
the platform has to involve two extra converts (as needed):
organizing discrete blocks to a contiguous buffer at sender 
and restoring discrete blocks from bytes at receiver.
However, due to limited HBM,
the sender may fail to prepare the contiguous buffer for all pending prompts.
Further, there exists a conflict between
1) block-free transfer for entire KVCache (all layers) with transparency to services
and 2) per layer triggers with less transfer time but revision required on models.
The tradeoff should be considered according to the model (whether to enable per layer transfer).
And, at last, the transfer should be stable in large scale (multiple hops involved).

%% file: tex/design.tex
\section{Design on Serving LLMs at Scale}
\label{sec:design}

\subsection{Overview}
\begin{figure}[!t]
    \centering
    \includegraphics[width=3.3in,height=1.861in]{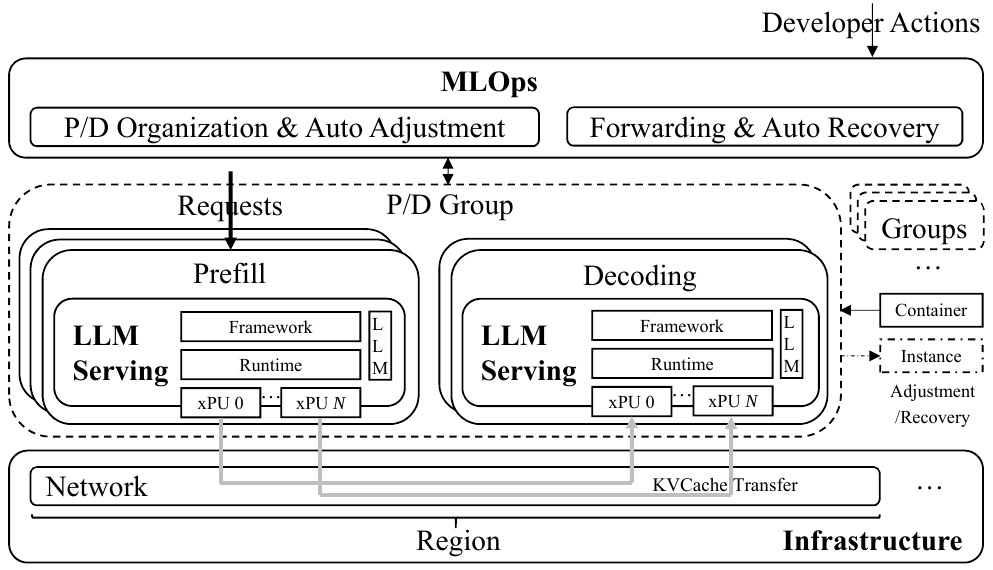}
    \vspace{-6pt}
    \caption{Overview of \sysname}
    \label{fig:design_architecture}
    \vspace{-11pt}
\end{figure}

\sysname is proposed as an end-to-end system,
complying with the paradigm of MLOps (machine learning operations),
serving disaggregated LLMs over tens of thousands of xPU devices,
as shown in Fig.~\ref{fig:design_architecture}.
\sysname contains three components:
1) the \textbf{MLOps}, 
i) manages the services (also the scenarios) and infrastructure 
(service-level resource management, e.g., inference at daytime and training at night),
ii) triggers the workflows of deployment and scaling for LLMs according to developer actions,
and iii) enables automatic request forwarding, traffic controls and recovery upon detected faults;
2) the \textbf{LLM Serving},
i) coordinates with MLOps for instance-level management (i.e., form a group of P/D instances),
ii) serves the prompts among P/D instances with all necessary controls and communications,
and iii) integrates the full stack for inference, 
including the model, the runtime (e.g., Ascend), and the framework (e.g., MindSpore);
and 3) the \textbf{infrastructure} 
contains tens of thousands of machines and xPU devices (e.g., NPUs), 
connected via the network (multiple regions),
with the basic ability of disaster recovery and load balance.

For the first challenge,
the LLM Serving coordinates with the MLOps in \sysname, 
mapping the service (also the scenarios) with RoCE as needed for fine-grained organization (P/D groups),
dynamically adjusting the ratios 
(minimize the mismatch regarding the processing capability),
and enabling the recovery with minimum cost upon auto fault detection,
as shown in subsections 
\S\ref{sec:design_roce}, \S\ref{sec:design_adjustment}, and \S\ref{sec:design_recovery}, respectively.

For the second challenge,
multiple gateways retry the requests among prefill instances within a time period to balance the workload,
in which the prefill may reject the requests if it is occupied.
If no prefill accepts the requests and its timeout threshold is reached,
\sysname terminates it without further forwarding.
That is, gateway helps the requests to find the idle prefill
(control the input traffic to match the processing capability for given P/D instances),
as illustrated in \S\ref{sec:design_pull}.

And for the third challenge,
\sysname enables D2D KVCache transfer in bytes,
where the sender prepares a contiguous buffer,
and the receiver recovers the blocks via RecvScatter,
as shown in \S\ref{sec:design_transfer}. 
Further details about infrastructure 
and configurations (switches and NPUs) 
are shown in \S\ref{sec:design_configuration}.

\subsection{P/D Organization upon RoCE}
\label{sec:design_roce}

\textbf{Map of Service with RoCE}:
The resources are managed at service granularity (coarse-grained),
in which the resources are assigned to inference at daytime and training at night,
and the service-level scaling copes with the changes on the traffic.
Unfortunately, coarse-grained management is insufficient.
Various scenarios, even in the same service, have quite different demands on scaling and adjustment,
which requires the platform to support the isolation of P/D instances among scenarios.
Thus, the need of organizing any instances to form a P/D group (fine-grained) is inevitable,
where such a group works for specific scenario (for related developers), is isolated to others,
and also obeys the management and scaling from superior service.
Due to fine-grained demands on scaling and adjustment,
for dynamic group organization,
a map between the service (and further scenarios) and P/D instances is a must.
That is, the relationship on P/D instances can be mapped to its relationship on RoCE.
Note that each prefill instance has the chance to forward the request (with KVCache)
to any decoding instance in a group for generating follow-up tokens.

Multiple xPU devices (e.g., NPUs) are assigned to one instance (i.e., via flexible container),
in which each device has the RoCE IP in the region (a cluster contains multiple regions for disaster recovery).
To form the relationship between P/D role and RoCE, we use the format like <P, \{<IP1, ...>, ...\}>,
where ``P'' refers to the prefill role, and each prefill instance in this group is described using its RoCE IPs.
For example, IP1 to IP8 refers to 8 xPU devices of an instance.
Since all the RoCE IPs are determined in advance via container,
organizing these information to form a logical group first,
and then triggering the initialization 
(establish connections and deploy the model) should be well orchestrated.

\begin{figure}[!t]
    \centering
    \includegraphics[width=3.3in,height=1.859in]{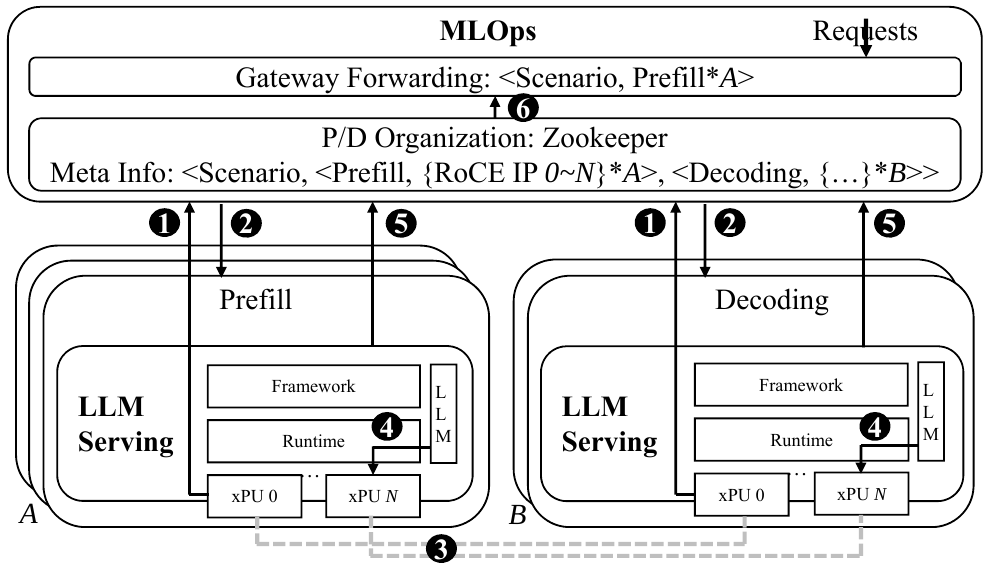}
    \vspace{-6pt}
    \caption{Workflow of P/D Setup for a Group}
    \label{fig:design_pd_setup_ingroup}
    \vspace{-8pt}
\end{figure}

\textbf{Workflow of P/D Setup in Groups}:
As shown in Fig.~\ref{fig:design_pd_setup_ingroup},
the workflow contains two parts: gathering information and initializing a group.
For those containers (triggered by Kubernetes~\cite{k8s}) to form a P/D group,
LLM Serving coordinates with MLOps to establish the relationship between the service (or scenario) and RoCE.
\protect\circled{1} LLM Serving (the resident process per instance) first obtains the RoCE IPs (e.g., via hccn\li tool).
Then, LLM Serving organizes the RoCE IPs in order, according to device IDs of assigned xPUs.
At last, LLM Serving reports to the Zookeeper~\cite{zookeeper} of MLOps,
in which all RoCE IPs are collected and recorded to target service (or scenario).
The Zookeeper completes the information gathering until the number of reports match the instance number.
If failures occur during the collection,
MLOps retries within pre-defined time threshold.
\protect\circled{2} As long as the Zookeeper completes the collection,
initialization order is delivered.

After receiving initialization order,
\protect\circled{3} LLM Serving establishes the connections (with verifications) 
and \protect\circled{4} deploys pre-compiled models.
The establishment continuously waits all possible connections until a timeout occurs.
After establishment,
all instances individually load pre-compiled model from a file system (e.g., scalable file service SFS).
The models loaded by prefill and decoding instances are different.
After initialization,
\protect\circled{5} LLM Serving enables regular health reports to the Zookeeper 
(besides the first, tens of seconds per follow-up reports).
\protect\circled{6} The whole workflow is completed until the Zookeeper confirms all reports from instances,
in which all prefill instances are labeled as the entrance for requests.

\textbf{Pre-compiled Model Loaded in Minutes}:
To avoid the waste on compilation,
the models for both prefill and decoding are pre-compiled, 
via subsequent task after training,
and stored to a file service (for loading from any instance).
LLM with hundreds of billion parameters is loaded within minutes.

\subsection{P/D Adjustment with MLOps}
\label{sec:design_adjustment}

\textbf{Group-based Scaling}:
The scaling covers a wide range of demands.
For the demands with P/D ratio unchanged
(e.g., the traffic surges, but the combination of requests does not change),
the scaling is conducted upon groups.
MLOps provides friendly UI for developers to scale-out/in P/D instances by groups (manually),
and auto scale-out/in via a time trigger.
The scale-in involves removing the map of the service (or scenario) and RoCE in the Zookeeper of MLOps,
to prevent the traffic from removed groups (further actions on non-interrupting the service shown later).
All data in the instances from removed groups are then erased, and the instances would be released.
Note that the resources are charged according to the duration of usage.
The scale-out involves using more containers (without any states) to form P/D groups as needed.
The process obeys the workflow of P/D setup mentioned before.
We should mention here that,
the workflow of P/D setup assumes the containers are stateless,
to facilitate the resource relocation among scenarios or even among services.

For the demands of rolling upgrades (e.g., model update via LoRA, prompt via prompt engineering, etc.),
the upgrade is performed based on the origin groups with unchanged P/D ratio,
and then further changes on the P/D ratio (shown later) as needed.
Since the upgrade is performed one group after another,
the upgrade actually does not involve the service interruption.
Note that each group receives a proportion of traffic for inference (at most group-level failure).

\begin{figure}[!t]
    \centering
    \includegraphics[width=3.3in,height=1.859in]{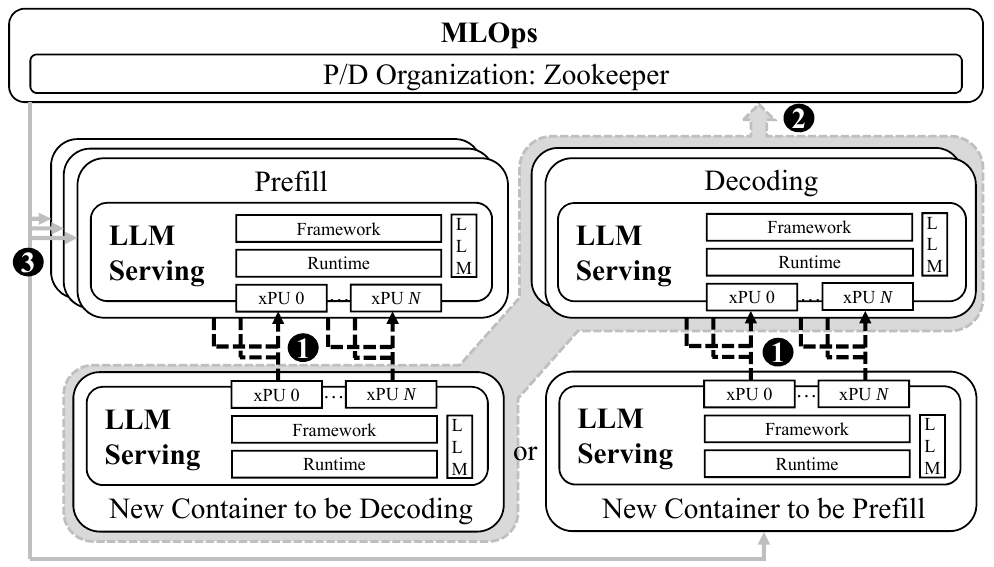}
    \vspace{-7pt}
    \caption{Dynamic RoCE for P/D Adjustment}
    \label{fig:design_dynamic_roce}
    \vspace{-12pt}
\end{figure}

\textbf{Dynamic RoCE Construction}:
Dynamically changing the P/D ratio without the service interruption facilitates 
the adjustments for both scaling and efficiency demands,
which essentially requires RoCE construction (and re-construction) for newly added but stateless containers.
As shown in Fig.~\ref{fig:design_dynamic_roce},
dynamically changing the P/D ratio contains two steps:
RoCE construction for newly added containers (stateless and no P/D role) and taking effect of new P/D ratio.
Note that removing existing instances is implemented via similar two steps.

Integrating newly added but stateless containers into existing RoCE involve new connections
between these containers with existing P/D instances.
More specifically,
the Zookeeper sends existing RoCE map recorded to newly added containers.
\protect\circled{1} The LLM Serving then triggers new connections from these containers,
in which the RoCE IPs of newly added containers are gathered and sent to existing P/D instances.
After receiving all the confirmations,
newly added containers can communicate with existing prefill or decoding according to the role.
Then, similar to the workflow of P/D setup,
these containers load pre-compiled model also based on the role,
and \protect\circled{2} send health reports to the Zookeeper.
After receiving the reports,
\protect\circled{3} the Zookeeper updates all meta information of decoding 
to prefill instances (also contains newly added ones),
to facilitate further forwarding from prefill to decoding.

\textbf{Ratio Adjustment for Changes}:
As mentioned before, the rolling upgrades are performed based on unchanged P/D ratio first.
Then, via profiling in advance,
or triggered by developers manually,
or monitoring latency changes (shown later),
the adjustment on P/D ratio is adopted.
The demands of P/D ratio adjustment derive from both
content change (i.e., prompt engineering, with the changes on prompts and tokens generated) 
and traffic change (i.e., the combination of requests changes).
If the profiling is performed,
the desired P/D ratio (shown later) is determined.
Then, related groups are adapted to such desired P/D ratio using dynamic RoCE construction
(gradually update those groups).
If the overall number of instances is fixed for a specific scenario,
re-construction of some groups is necessary 
(gradually release part of the instances and add them to the other groups to from new P/D ratio).
Otherwise, the adjustments may involve new instances
(for intermediate steps, and may release redundant ones after adjustments).

\begin{figure}[!t]
    \centering
    \includegraphics[width=3.3in,height=1.859in]{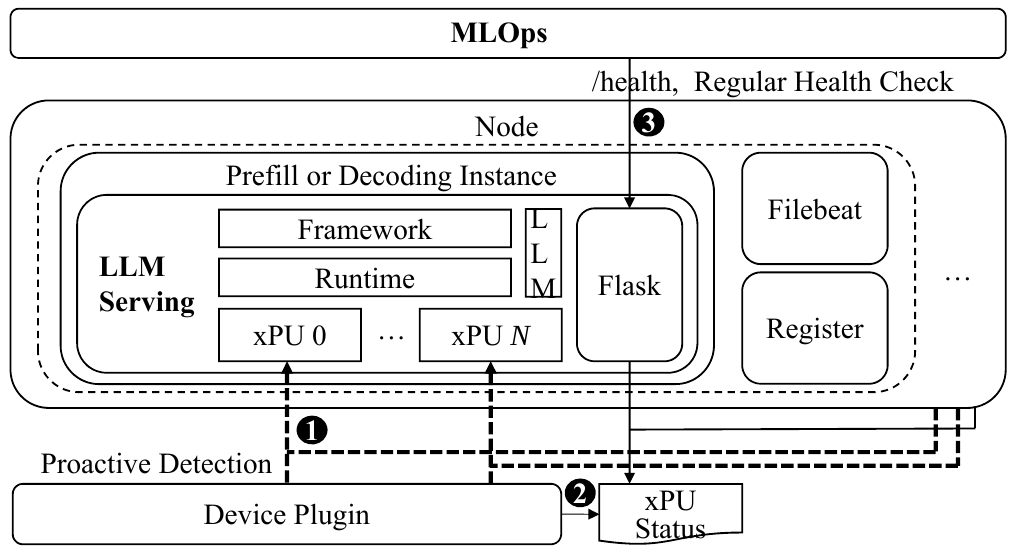}
    \vspace{-4pt}
    \caption{Automatic Fault Detection}
    \label{fig:design_auto_detection}
    \vspace{-8pt}
\end{figure}

The monitor records the averaged E2E latency and TTFT (actually $T_{p}$, here for convenience) per scenario.
Once the proportion of TTFT and the averaged E2E latency change dramatically 
(for both content change or traffic change),
the adjustment on P/D ratio is recommended.
For example, with the increase on both E2E latency and the proportion of TTFT,
more prefill instances are needed.
In contrast, with the increase on E2E latency but decrease on the proportion of TTFT,
more decoding instances are needed.
Profiling a certain pattern on prompts benefits determining the optimized P/D ratio:
\begin{gather}
	n_{p}b_{p}\cdot 1/T_{p} \;\approx\; n_{d}b_{d}\cdot 1/T_{d}, \label{minimize_mismatch}
\end{gather}
which essentially minimizes the mismatch between prefill and decoding regarding the processing capability
(single point failure should be also avoided per scenario).
Profiling involves extra resources in advance.
An adaptive approach is to gradually increase the instances,
upon both TTFT proportion and increased E2E latency,
catering to the changes over time.

\subsection{Minimum-cost Auto Recovery}
\label{sec:design_recovery}

\textbf{Automatic Fault Detection}:
Various faults occur during LLM inference, especially when disaggregated paradigm is adopted.
The most common phenomenon is timeout.
Timeout actually helps the early intervention (either in instance or at gateway) for quick response.
Blindly re-try or re-start fails to dig out the real reasons behind timeouts,
which leaves the uncertainty to the whole system.
Note that xPUs are continuously used all day long for both training and inference (auto switch for tidal traffic).
As shown in~\cite{huggingface_bloom}, 
about 1 or 2 faults occur per week over the cluster with 400 GPUs.
With the growth of xPU device number (i.e., tens of thousands of xPUs), 
the faults are very common (both recoverable and unrecoverable).
As in Fig.~\ref{fig:design_auto_detection}, for Ascend NPUs,
a customized container (with resident process) upon Ascend Device Plugin~\cite{ascend_device_plugin}
is deployed per node (e.g., a physical machine, may contain multiple instances).
\protect\circled{1} The resident process regularly detects the faults 
and \protect\circled{2} records the results (xPU status) to a file.
Such file is mounted to all instances per node.
Further, the faults are classified into multiple levels,
in which some are recoverable without node-level recovery.
\protect\circled{3} MLOps regular checks the xPU status via a Flask request and triggers auto substitute.

\textbf{Recovery without Interruption}:
If one instance in a group is detected as failure,
MLOps triggers auto recovery by substituting it with a new instance.
More specifically,
after the detection, the meta information recorded in the Zookeeper is updated (logically removed),
to avoid forwarding further requests.
Then, the meta information is sent to all instances in this group 
to avoid actual transmission and forwarding.
Afterwards, like ratio adjustment, 
a newly added but stateless container is involved.
After dynamic RoCE construction, loading pre-compiled model, and reporting health to the Zookeeper,
the Zookeeper updates the meta information of this container to existing instances
(only updates decoding information in prefill).
Then, all the status of fault one would be erased.
For running requests, the protection shown later ensures the termination.
No further requests are forwarded to the substitute ensured by the Zookeeper (no harm to others).
Note that one newly added container is involved (minimum).

\textbf{Protection over faults}:
Except for xPU devices,
the faults also occur during calculation, transmission, or even forwarding.
Each component in the system has the chance to involve the faults.
Then, the protection like early intervention is needed,
as early as possible for the completeness.
For example, if the fault occurs in a decoding instance,
related prefill instances (LLM Serving is aware of the timeout) and MLOps perform all cleanup actions
(not consider request migration in this paper due to maintained streaming),
like stopping the connections, 
responding with users using default texts (or retrieval-augmented generation, RAG~\cite{fan2024surveyrag}),
and updating the meta information of decoding to avoid further forwarding, etc.
Such protection (for the completeness) like stopping and responding is necessary.
Otherwise, the zombie connections easily involves harms and uncertainty.

\subsection{On-demand Forwarding for Idle Prefill}
\label{sec:design_pull}

\textbf{Streaming Responses via SSE}:
The autoregressive LLM generates tokens one after another.
To improve the user experience, it is preferred to deliver the token 
(converts it to the texts in advance) to the user as long as it is generated.
Such streaming responses are implemented via server-sent events (SSE), 
which is essentially unidirectional,
enabling the client to receive automatic updates from the server using the HTTP connection.
Note that the entire network should support such feature 
(i.e., decoding -> prefill -> gateway).
Then, each component should maintain such connections for SSE.
As for each prefill instance, it only maintains the requests forwarded to related group.
However, for the gateway,
it maintains extensive connections per second.
The gateway just forwards the streaming responses instead of complex resolution.
And the scheduler is integrated with the gateway to avoid further forwarding.
Actually, there are multiple gateways in a cluster.

\begin{figure}[!t]
    \centering
    \includegraphics[width=3.3in,height=1.859in]{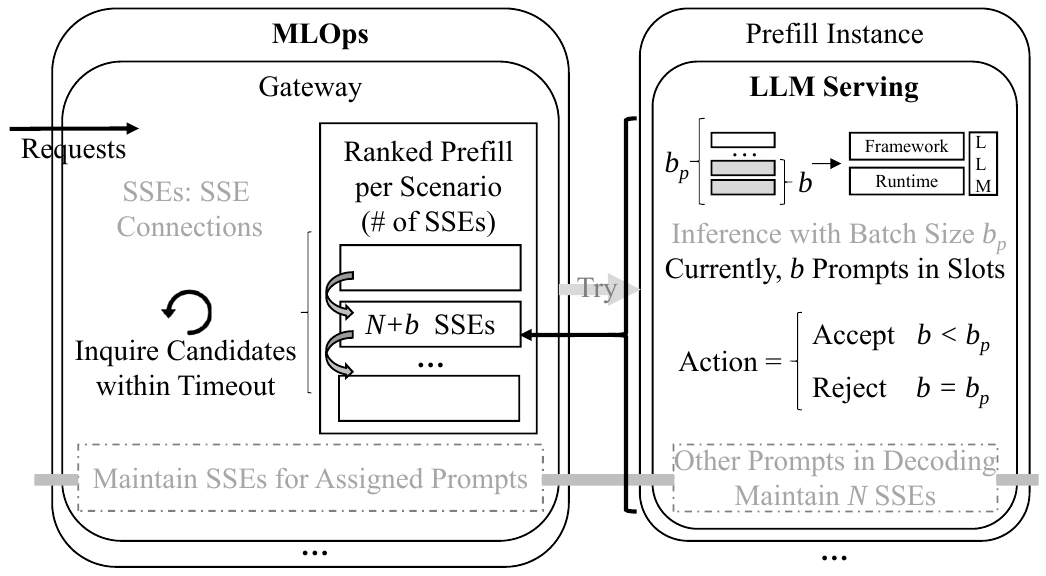}
    \vspace{-6pt}
    \caption{On-demand Forwarding for Idle Prefill}
    \label{fig:design_forwarding}
    \vspace{-8pt}
\end{figure}

\textbf{Idle Prefill or Rejection}:
The prompts already assigned in the queue fail to be re-directed to other idle prefill instances.
And, the prediction upon the queue status (pending tokens) 
is inaccurate due to batch processing and prefix-aware KVCache.
To avoid local-optimum and also to avoid preemptive scheduling,
the local queue in prefill is removed.
Actually, the SSE directly hints the workload of a group at gateway,
since the connections are maintained for streaming responses.
However, SSE covers the entire LLM lifecycle 
(also contains token generation in decoding).
The number of SSE connections alone is inadequate 
(i.e., cannot distinguish whether prefill is completed).
Therefore, the prefill is enabled with the ability of accepting or rejecting the requests.
If the prefill is idle, it accepts new requests for batch processing.
Otherwise, the rejections are sent to the gateway.
Note that, a prompt continuously occupies one slot in prefill 
if it is waiting for KVCache transfer to a decoding for follow-up generation.

\textbf{On-demand Forwarding at Gateway}:
Each one of the gateways performs on-demand forwarding based on the number of connections and prefill feedbacks,
as shown in Fig.~\ref{fig:design_forwarding}.
A request is assigned to a specific prefill by gateway,
in which the gateway chooses the one with the least number of SSE connections.
Note that the prefill instances are organized in order at the gateway.
At that moment, the gateway supposes this prefill is more likely to be idle.
Here, the worst case is: the desired prefill rejects the request, implying it is occupied.
Then, the gateway consider other candidates 
(e.g., a subset of prefill instances top ranked) within a time period (e.g., the timeout threshold).
The gateway inquires these prefill instances one after another.
Although several retries are performed,
the acceptance implies the request must be assigned to an idle prefill
(i.e., waiting at the gateway instead of the local queue belonging to a prefill).
The gateway achieves
\begin{gather}
	I_{t} \;\approx\; n_{p}b_{p}\cdot 1/T_{p}, \label{input_traffic}
\end{gather}
where the requests assigned matches the processing capability of prefill per scenario.
To facilitate full batch processing,
the gateway continuously forwards the requests to one idle prefill until it is busy
(or using chunked prefill for continuous batching).
The request terminates (early intervention),
if none prefill is idle and the waiting time exceeds the threshold.

\begin{figure}[!t]
    \centering
    \includegraphics[width=3.3in,height=1.859in]{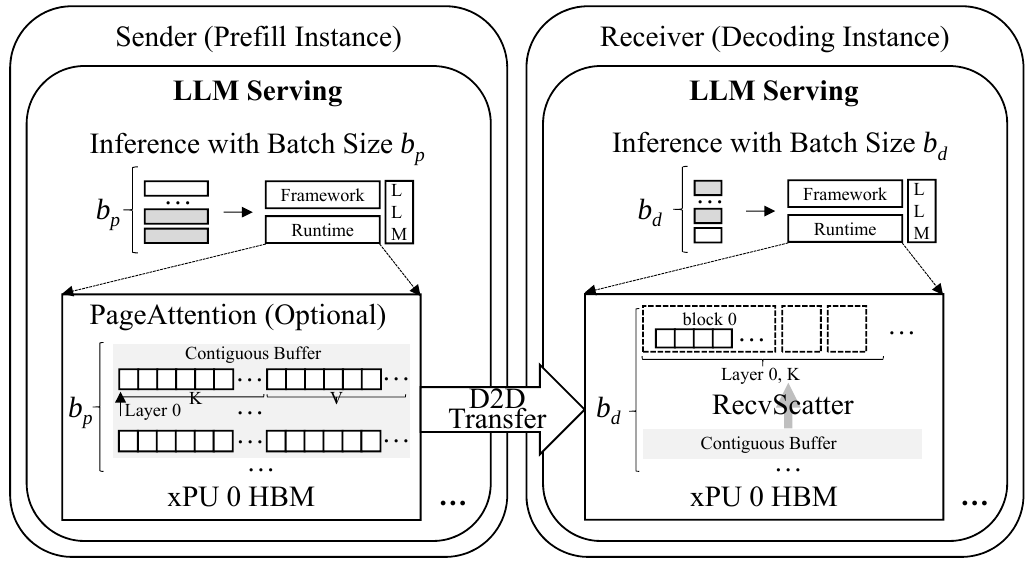}
    \vspace{-7pt}
    \caption{Block-free D2D KVCache Transfer}
    \label{fig:design_block_free_transfer}
    \vspace{-9pt}
\end{figure}

\subsection{Block-free D2D KVCache Transfer}
\label{sec:design_transfer}

\textbf{Contiguous Buffer at Sender}:
The desired KVCache transfer prefers to treat all the data as a whole in bytes.
Therefore, for key-value pairs (\textbf{K} and \textbf{V}) generated per layer,
the sender organizes the KVCache together 
in a contiguous buffer to facilitate the transfer.
As illustrated in Fig.~\ref{fig:design_block_free_transfer},
there are no discrete blocks at the sender,
and all the key-value pairs are managed one after another in the buffer.
Given the index of a layer, the offset and the length can be quickly calculated 
according to the prompt length and the hidden size of a model (also the precision of floats).
Either the transfer per layer or the entire model can be triggered 
by using different offsets and lengths.
If there are multiple prompts,
it is hard to ensure the prepare of contiguous buffers for all of them.
However, benefit from fine-grained P/D organization (i.e., homologous prompts per scenario)
and on-demand forwarding upon rejections (i.e., limited prompts entered),
reserving all of these contiguous buffers for KVCache transfer is possible in prefill in advance.
Before each transfer for a specific range of key-value pairs in the buffer,
one communication with a low cost exchange of the meta is necessary between sender and receiver.

\textbf{RecvScatter at Receiver}:
As shown in Fig.~\ref{fig:design_block_free_transfer},
the RecvScatter refers to either a function or an operator, 
to facilitate the tradeoff between transparency and flexibility.
As mentioned before,
a conflict exists:
1) block-free transfer for entire KVCache (all layers) with no revision on models (transparency to services);
2) per layer triggers with less transfer time but revision required on models (more flexibility).
Note that the operator implementation does not interrupt the computation of other operators in the stream.
It is preferred to trigger either function or operator implementation
according to the scenario (per layer transfer depends on explicitly model revision).
\sysname facilitates such revision during or after the training 
(manually insert operators or during model pre-compilation via convert tasks).
Whether to use per layer transfer also affects when to select an idle decoding
(before or after prefill).

\textbf{Asynchronous Retrieval}:
The decoding has limited HBM space for KVCache transfer.
And, the KVCache transfer and the decoding inference can be concurrent (i.e., asynchronous retrieval).
By default, one KVCache is transferred and added to running batch after a completed request (continuous batching).
Note that several concurrent transmissions share the D2D bandwidth,
whose average completion time is actually less than the transfer one after another.
Further, similar to sub-optimum in prefill scheduling, 
the space left for KVCache transfer actually forms a local queue.
Asynchronous retrieval is preferred, but just waiting in such queue sacrifices other decoding choices.
Thus, the capacity of such queue is relatively small,
to facilitate on-demand use 
(i.e., a completed request triggers next retrieval).
The retrieval here also does no harm to the running requests in decoding.
After asynchronous retrieval,
the pending KVCache occupies the slot (belonging to the request just completed)
and is valid in the next iteration (decoding iteratively generates tokens for batch requests).

\begin{figure}[!t]
    \centering
    \includegraphics[width=3.2in,height=1.802in]{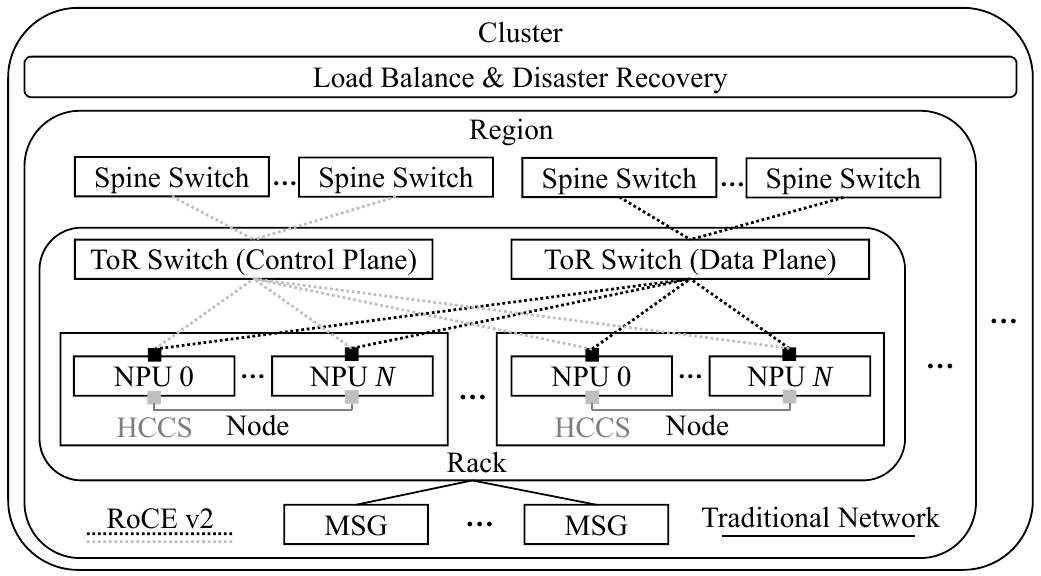}
    \vspace{-5pt}
    \caption{Infrastructure contains tens of thousands of xPUs.}
    \label{fig:design_infrastructure}
    \vspace{-7pt}
\end{figure}

\subsection{Infrastructure and Configurations}
\label{sec:design_configuration}

\textbf{Network Organization}:
As illustrated in~\ref{fig:design_infrastructure},
there are multiple regions in a cluster,
in which the resources are managed according to the computing capability of xPU devices (i.e., NPUs).
Each region contains thousands of NPUs and related resources.
Actually, a rack contains multiple Atlas servers,
in which each server (node) contains multiple NPUs.
The HBMs of NPUs are tens of GBs,
and the NPUs belonging to the same node are connected via HCCS~\cite{hccs}.
Moreover, the NPUs are directly connected to top-of-rack (ToR) switches with RoCE v2 enabled.
Multiple ToR switches are prepared in a rack for both control plane and data plane.
The spine switches are further connected to ToR switches for cluster-level transfer.

The requests are forwarded via 
elastic load balancer (ELB), software load balancer (SLB) and model service gateway (MSG),
where these three components work together for 
load balance, traffic control and forwarding the requests to specific services 
(to each prefill per scenario for disaggregated LLM).
In this paper, we only consider the strategy at MSG gateway.

\textbf{Configurations in Large Scale}:
The NPUs are directly connected to ToR switches with RoCE v2 enabled.
Therefore, the transfer does not need the host network (one hop less),
which requires the meta for RoCE connections being maintained in HBM.
Due to limited HBM,
the space left for meta information should be well orchestrated.
Benefit from fine-grained P/D organization and 
the transfer between the NPUs labeled as the same index among instances,
the space left is acceptable (i.e., hundreds of MB).
Further compression is performed for supporting more NPUs (for dynamic RoCE construction).
Moreover, each D2D KVCache transfer incurs multiple sub-transfers 
between a pair of P/D instances (i.e., due to multiple xPUs).
The conflict should be avoided among those sub-transfers,
which requires the infrastructure to fully utilize the path diversity between ToR and spine switches.
Note that a conflict may lead to hundreds of milliseconds.

\textbf{Disaster Recovery over Regions}:
Multiple regions are used in a cluster,
in which the devices and related resources are divided among those regions.
The P/D instances are organized via groups (fine-grained manner) per scenario and can be deployed to any region
(each one with suitable NPU types is equipped with the infrastructure for KVCache transfer).
The ELB and SLB are responsible for load balance for scenarios.
Once the region-level failures occur,
other regions continuously serve the requests without service interruption.

%% file: tex/evaluation.tex
\section{Performance Evaluation}
\label{sec:evaluation}

\begin{figure*}[t]
    \begin{subfigure}[t]{0.22\textwidth}
        \setlength{\abovecaptionskip}{2pt}
        \includegraphics[width=1.62in,height=1.265in]{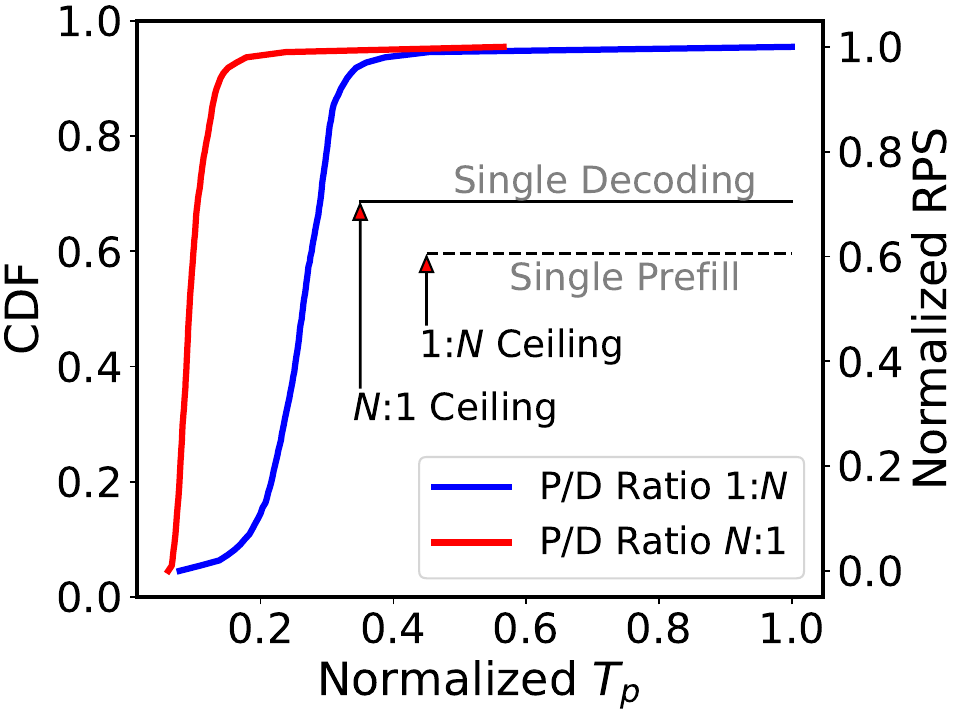}
        \caption{Mismatch and Bottleneck} 
        \label{fig:eval_1}
    \end{subfigure}
    \hfill
    \begin{subfigure}[t]{0.22\textwidth}
        \setlength{\abovecaptionskip}{2pt}
        \includegraphics[width=1.62in,height=1.265in]{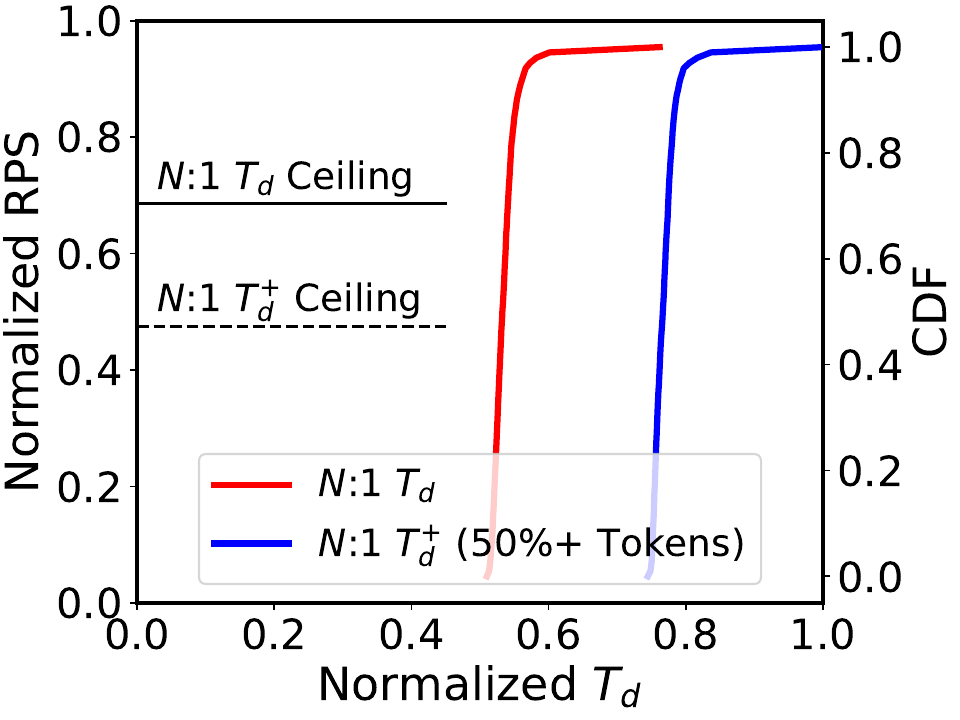}
        \caption{$T_{d}$ Affects Capability}
        \label{fig:eval_2}
    \end{subfigure}
    \hfill
    \begin{subfigure}[t]{0.22\textwidth}
        \setlength{\abovecaptionskip}{2pt}
        \includegraphics[width=1.62in,height=1.265in]{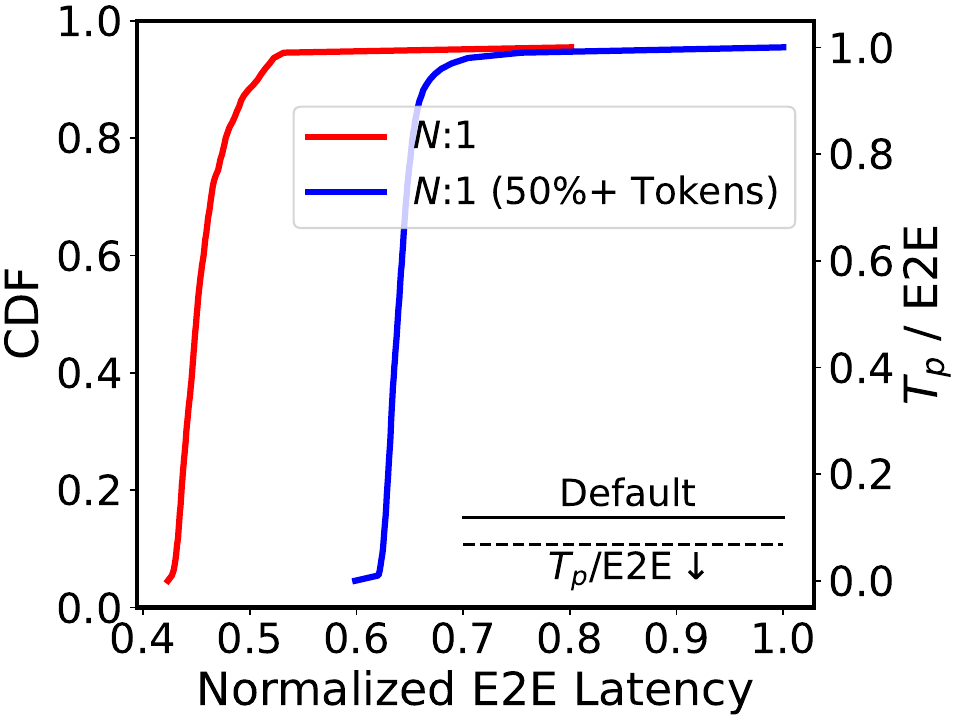}
        \caption{Proportion of $T_{p}$/E2E} 
        \label{fig:eval_3}
    \end{subfigure}
    \hfill
    \begin{subfigure}[t]{0.22\textwidth}
        \setlength{\abovecaptionskip}{2pt}
        \includegraphics[width=1.62in,height=1.265in]{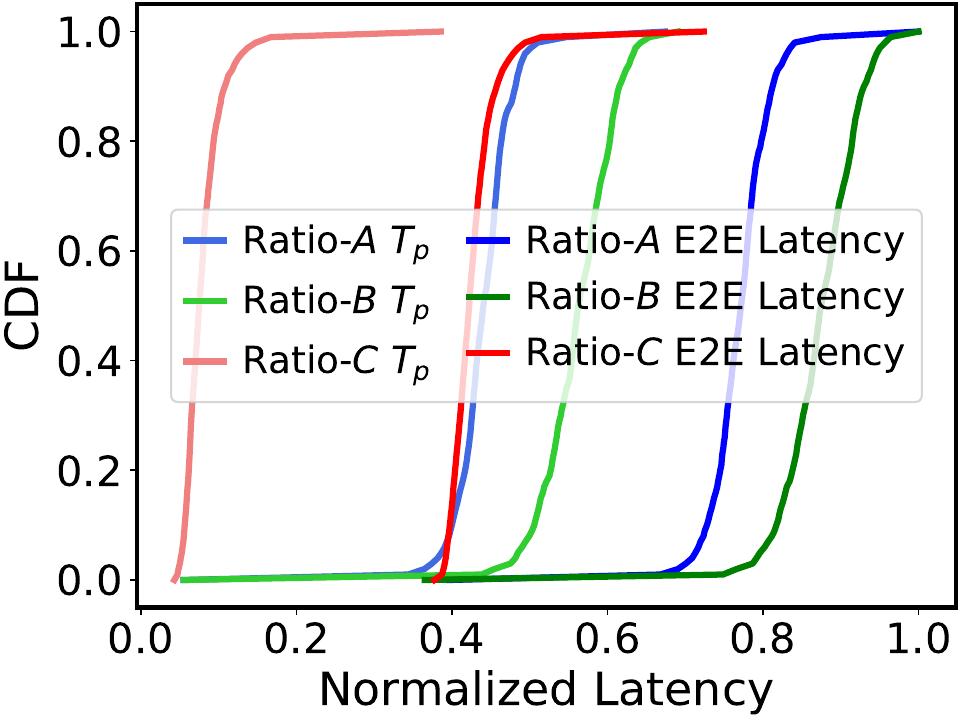}
        \caption{E2E under Various Ratios} 
        \label{fig:eval_4}
    \end{subfigure}
    \vspace{-7pt}
    \caption{Evaluation on \sysname: P/D mismatch incurs bottleneck; P/D ratios is adjusted for better performance.}
    \vspace{-1pt}
\end{figure*}

\begin{figure*}[t]
    \begin{subfigure}[t]{0.22\textwidth}
        \setlength{\abovecaptionskip}{2pt}
        \includegraphics[width=1.62in,height=1.265in]{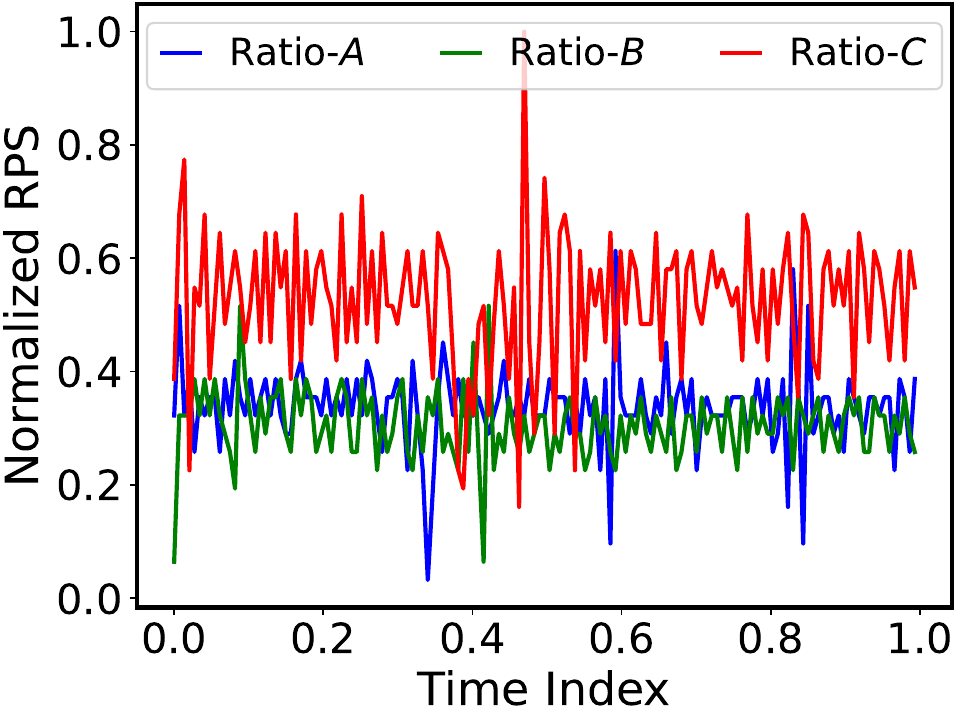}
        \caption{Throughput under Ratios} 
        \label{fig:eval_5}
    \end{subfigure}
    \hfill
    \begin{subfigure}[t]{0.22\textwidth}
        \setlength{\abovecaptionskip}{2pt}
        \includegraphics[width=1.62in,height=1.265in]{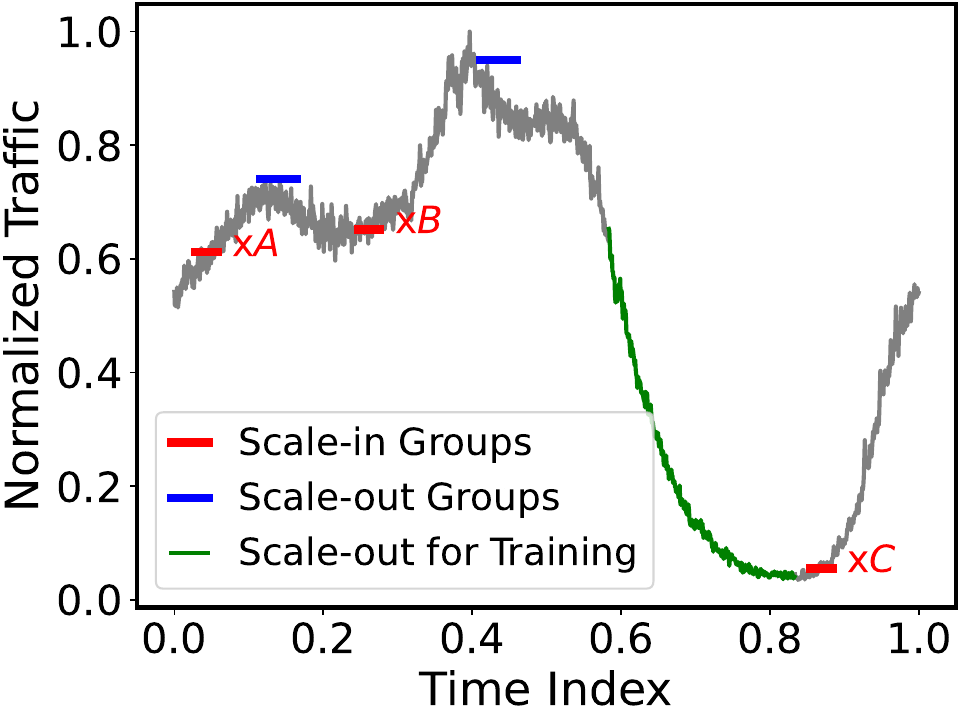}
        \caption{Auto Scaling upon Groups}
        \label{fig:eval_6}
    \end{subfigure}
    \hfill
    \begin{subfigure}[t]{0.22\textwidth}
        \setlength{\abovecaptionskip}{2pt}
        \includegraphics[width=1.62in,height=1.265in]{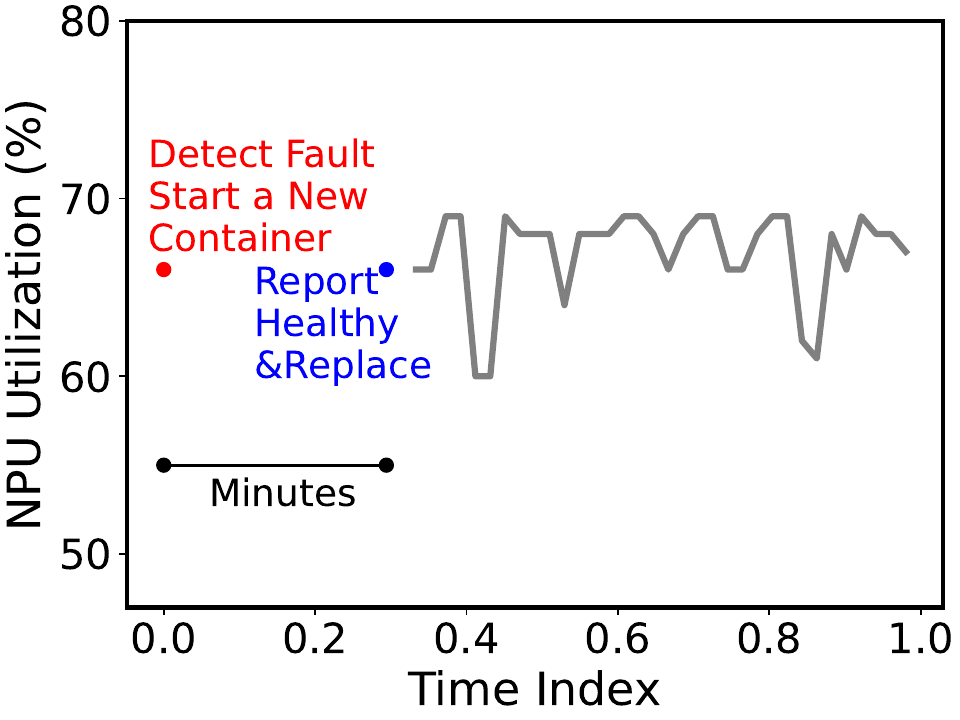}
        \caption{Recovery via a Substitute} 
        \label{fig:eval_7}
    \end{subfigure}
    \hfill
    \begin{subfigure}[t]{0.22\textwidth}
        \setlength{\abovecaptionskip}{2pt}
        \includegraphics[width=1.62in,height=1.265in]{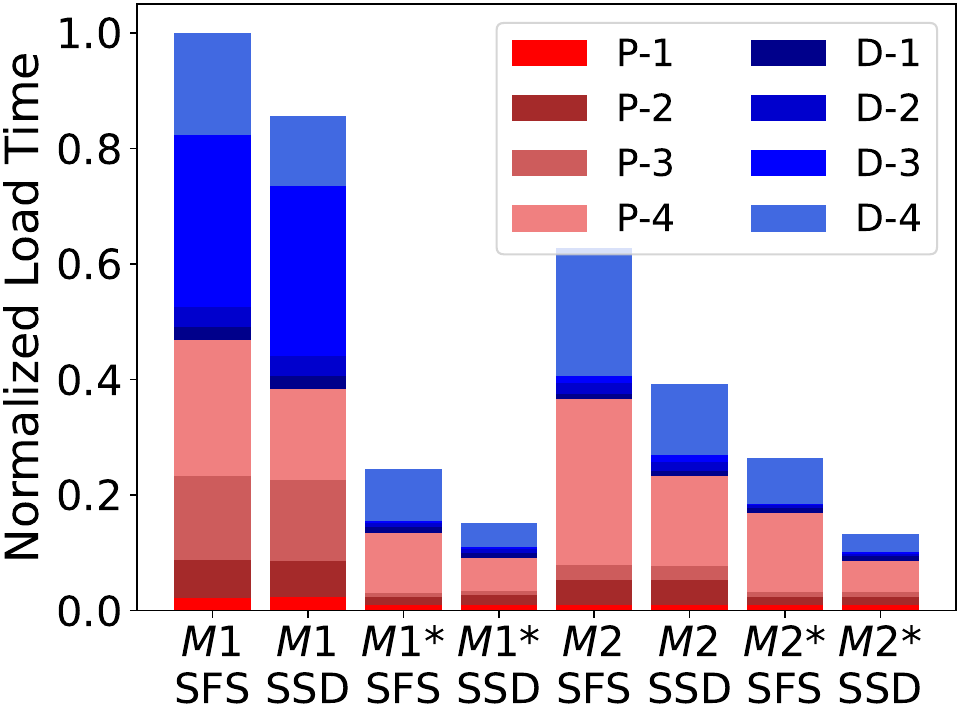}
        \caption{Loading Compiled Models} 
        \label{fig:eval_8}
    \end{subfigure}
    \vspace{-7pt}
    \caption{Evaluation on \sysname: Both P/D adjustment and P/D organization are adopted.}
    \vspace{-8pt}
\end{figure*}

\subsection{Experimental Setup}
\textbf{Workloads}:
All the requests analyzed are derived from real services (fine-tune datasets), 
instead of the production cluster, due to the privacy policy.
After the collection, those requests are further classified, 
to verify the similarity on P/D behaviors (already shown before).
We should mention here that
the requests from upstream services actually contain the scenario information (labelled after the intention understanding),
which helps \sysname to perform the fine-grained P/D organization and the forwarding upon the similarity (prompt length or scenario).
Except for the content analysis,
the system metrics are measured in both mirror environment and the production cluster.
The machine types (multiple choices equipped with NPUs) used in mirror environment (hundreds of NPUs) 
are actually a subset of that in production cluster (tens of thousands of NPUs).
Either SFS or SSD is used for storing compiled models.
The models used are the Pangu~\cite{ren2023pangusigma} (\sysname also supports other mainstream models), 
whose parameters also vary according to the scenarios.
All the results are normalized to a standard range 0$\sim$1 for illustration.
All the features are first implemented and evaluated in mirror environment and then deployed over all NPUs.

\textbf{Baseline}:
As mentioned in previous works,
disaggregated LLMs earn a significant improvement on both latency and throughput.
In this paper, we use the first commercial version for disaggregated LLMs as the baseline 
(overall 6.7x increase on throughput, compared with serving aggregated LLMs).

\subsection{E2E Performance}
\textbf{P/D Mismatch}:
The mismatch derives from the difference on processing capability between prefill and decoding.
As shown in Fig.~\ref{fig:eval_1},
$T_{p}$ (TTFT upon batch processing and prefixes cached) is quite different when using P/D ratios 1:$N$ and $N$:1.
Here, $N$ is the number of the instances larger than 1.
Fig.~\ref{fig:eval_1} further shows the processing capability of each single P/D instance (already normalized).
Then, blindly increasing the number of the instances for only one role is insufficient.
For example, the processing capability of one prefill instance is 0.6 in this scenario 
while that of decoding is about 0.7.
Blindly increasing the prefill instances actually enhances the prefill capability (0.6*$N$), but its bottleneck is still the decoding (0.7).
In contrast, blindly increasing the decoding fails to remove the bottleneck in prefill.
As in the modeling,
both $T_{p}$ and $T_{d}$ affect the processing capability
(i.e., with the growth of $T_{p}$ and $T_{d}$, the throughput drops).
As shown in Fig.~\ref{fig:eval_2},
given P/D ratio $N$:1 (this case generates less tokens),
with the increase of tokens generated,
the latency of decoding phase (labeled as $T_{d}^{+}$, 50\%+ increase) increases.
Furthermore, Fig.~\ref{fig:eval_2} shows the capability of decoding.
Compared with $T_{d}$,
increasing $T_{d}^{+}$ slows down the processing capability of decoding (although the bottleneck is still prefill).
As shown in Equation (\ref{minimize_mismatch}),
based on previous profiling,
minimizing the mismatch is necessary.
Except for profiling in advance,
dynamic bottleneck detection is enabled.
As shown in Fig.~\ref{fig:eval_3},
increased E2E potentially gives an alarm while the proportion of $T_{p}$ hints the P/D bottleneck.
With the growth of tokens generated and unchanged P/D ratio,
E2E latency increases,
and the proportion of $T_{p}$/E2E decreases,
which implies the decoding phase occupies much, and more instances are needed (motivating ratio adjustment).

\textbf{P/D Adjustment}:
By adopting P/D ratio adjustment (either profiling in advance or detected online),
the P/D mismatch actually decreases,
leading to improved latency and throughput.
Note that blindly increasing the input traffic, ignoring the bottleneck, easily leads to timeouts
(shown in Equation (\ref{input_traffic}), further solved using on-demand forwarding).
We should mention here that,
the tests are conducted by maintaining the constant requests (one completed triggers new one added).
And, the number of requests continuously increases until the success rate drops 
(E2E SLO, less than 100\%, i.e., no timeouts).
As shown in Fig.~\ref{fig:eval_4},
by using different P/D ratios,
$T_{p}$ and E2E are quite different (minimum is preferred).
Actually, the optimum ratio is obtained by using Equation (\ref{minimize_mismatch}).
Further throughput results are shown in Fig.~\ref{fig:eval_5},
under optimum P/D ratio,
the throughput overcomes the others by at least 60\%.
As for the latency,
minimizing the mismatch (in a group) actually decreases the probability of unnecessary waiting.
Note that, due to faster inference in prefill,
some KVCaches wait in prefill for idle decoding.
And for the throughput,
the resource utilization of P/D instances increases,
resulting in swallowing more requests without timeouts (fully utilize the pipeline).

\begin{figure*}[t]
    \begin{subfigure}[t]{0.22\textwidth}
        \setlength{\abovecaptionskip}{2pt}
        \includegraphics[width=1.62in,height=1.265in]{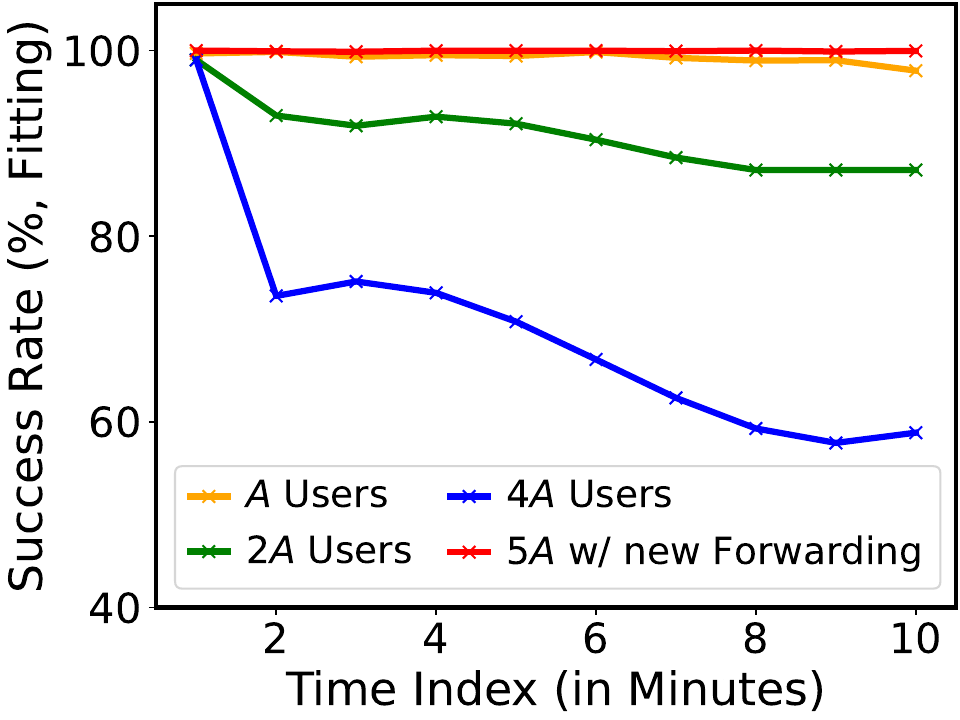}
        \caption{Changes on Success Rate} 
        \label{fig:eval_9}
    \end{subfigure}
    \hfill
    \begin{subfigure}[t]{0.22\textwidth}
        \setlength{\abovecaptionskip}{2pt}
        \includegraphics[width=1.62in,height=1.265in]{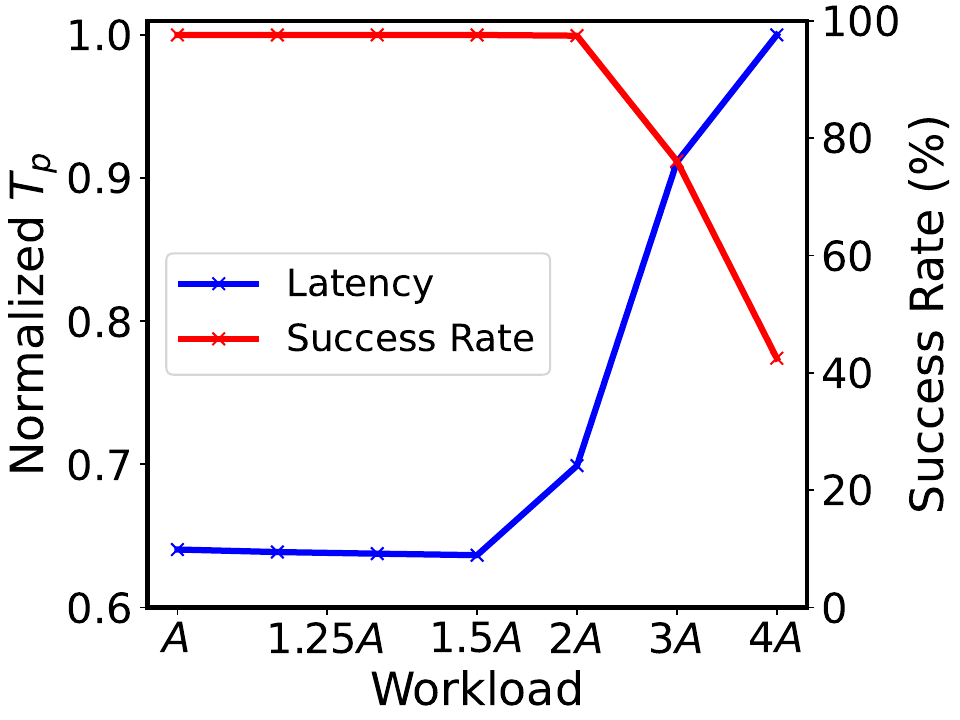}
        \caption{Success and Latency}
        \label{fig:eval_10}
    \end{subfigure}
    \hfill
    \begin{subfigure}[t]{0.22\textwidth}
        \setlength{\abovecaptionskip}{2pt}
        \includegraphics[width=1.62in,height=1.265in]{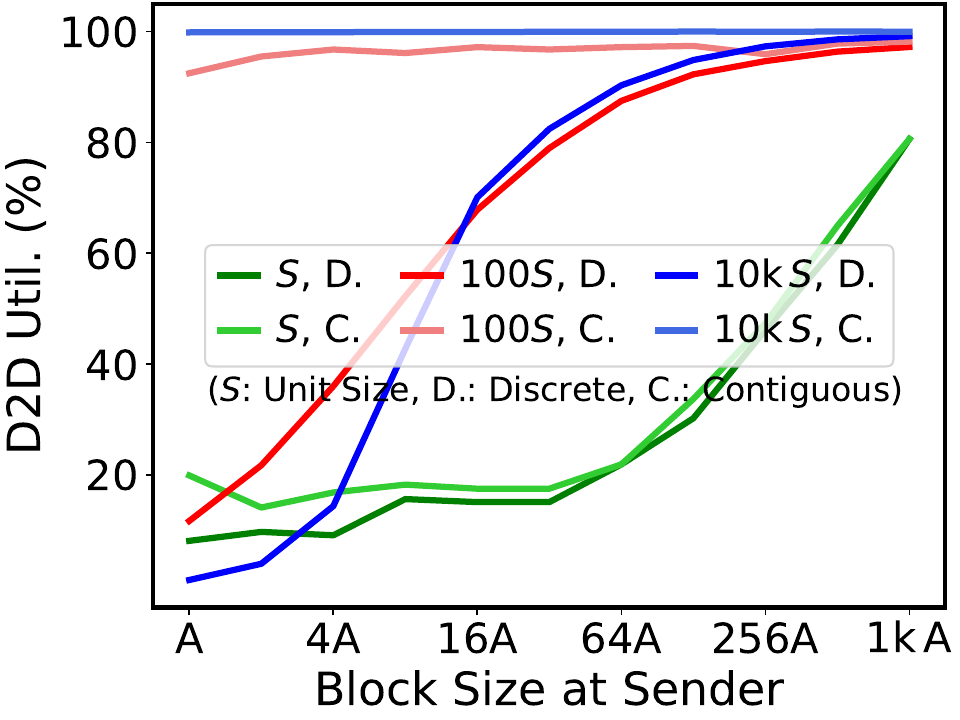}
        \caption{D2D Utilization} 
        \label{fig:eval_11}
    \end{subfigure}
    \hfill
    \begin{subfigure}[t]{0.22\textwidth}
        \setlength{\abovecaptionskip}{2pt}
        \includegraphics[width=1.62in,height=1.265in]{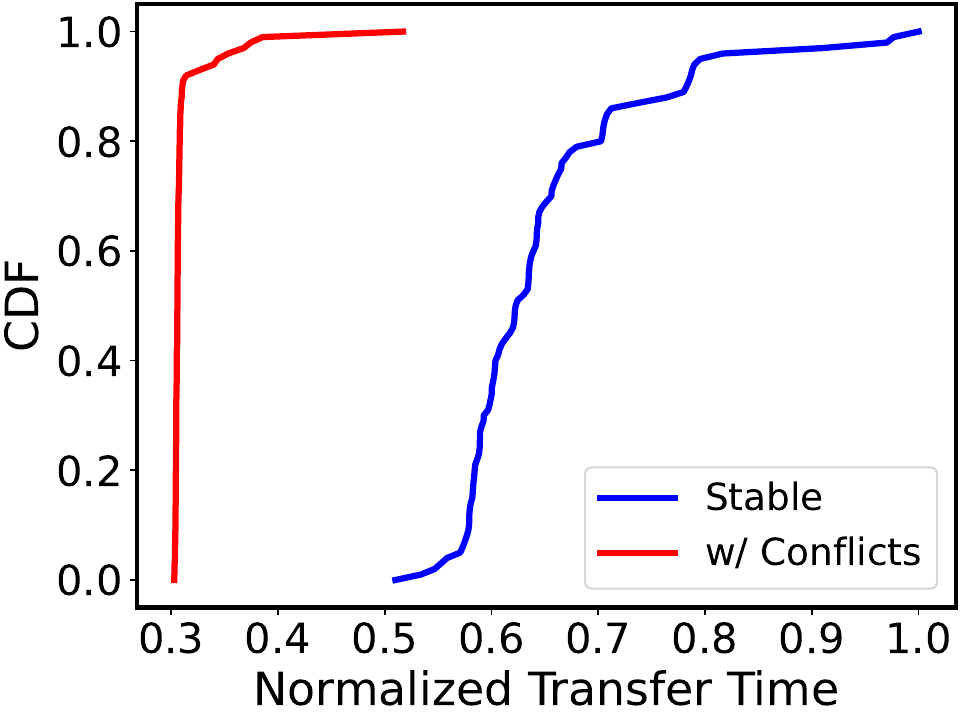}
        \caption{Stable Transfer Time} 
        \label{fig:eval_12}
    \end{subfigure}
    \vspace{-6pt}
    \caption{Evaluation on \sysname: Designed forwarding and transfer lead to higher success rate and lower D2D time.}
    \vspace{-5pt}
\end{figure*}

\textbf{Auto Workflows}:
MLOps facilitates auto workflows for both P/D organization and recovery.
As illustrated in Fig.~\ref{fig:eval_6},
the traffics are shown for a specific scenario during a whole day.
The green line shows the auto scaling between inference and training
(e.g., inference at daytime and training at night).
Furthermore,
the red lines show the scale-in actions by group granularity.
In this scenario,
there are three scale-in actions for catering to more traffic 
while two scale-out actions are performed.
The P/D organization,
including P/D setup, scaling (re-organization of a few groups among scenarios), and rolling upgrade, etc.,
is automatically performed based on the platform (P/D adjustment already shown before).
Equipped with auto fault detection,
the recovery is also conducted by using auto workflow.
As shown in Fig.~\ref{fig:eval_7},
After detecting the fault,
a new container is triggered and then as the substitute integrated into the P/D group.
After the integration,
stateless container acts as prefill or decoding role, 
and becomes a P/D instance.
The loading spends minutes, 
and after the substitute, the NPUs are occupied for inference. 
The loading is further illustrated in Fig.~\ref{fig:eval_8},
in which ``P'' and ``D'' refer to prefill and decoding, respectively (there are four further parts in loading).
Meanwhile, M1 and M2 are two different models,
and the symbol ``$*$'' refers to related optimizations.
Actually, SSD overcomes SFS during loading pre-compiled models.

\textbf{On-demand Forwarding}:
The original version uses the local queue in prefill,
and the gateway choses the one with minimum SSE connections.
As mentioned before,
already assigned prompts easily break timeout thresholds.
As shown in Fig.~\ref{fig:eval_9},
with the growth on workload (from $A$ users to 4$A$ users),
the success rate drops dramatically (from 100\% to the worst 57\%).
By removing prefill queues and invoking gateway retries,
with even heavier workload,
the success rate maintains at least 99\%.
Compared with the same workload,
the gap of success rate is up to 42.3\%.
Note that desired success rate is 100\%,
which implies no requests break the timeout thresholds and SLO is achieved.
The relationship between success rate and latency is shown in Fig.~\ref{fig:eval_10}.
Here, the timeout check is conducted before and after the prefill inference.
Thus, some prompts have already broken the timeout threshold during execution.
These prompts are still counted.
Although timeout intervention during prefill execution is useful,
it wastes the computing power of xPU and is actually ignored.

\textbf{Block-free Transfer}:
By using transferring KVCache as a whole and restoring the bytes to discrete blocks,
the utilization of D2D bandwidth improves, as illustrated in Fig.~\ref{fig:eval_11}.
Such gap derives from the bubbles,
in which frequent controls are involved for discrete block transfer, one after another.
The average transfer time decreases 46\%, compared with discrete block transfer.
As shown in Fig.~\ref{fig:eval_12},
with conflicts during multi-hop transfer,
the transfer time varies dramatically.

%% file: tex/related.tex
\section{Related Works}
\label{sec:related}

\subsection{Disaggregated LLM}
Numerous related references have emerged within a short six-month span: 
Splitwise~\cite{patel2023splitwise}, 
TetriInfer~\cite{cunchen2024tetriinfer}, 
DistServe~\cite{zhong2024distserve}, 
Dejavu~\cite{strati2024dejavu}, 
MemServe~\cite{hu2024memserve} 
and Mooncake~\cite{qin2024mooncake}. 
All prior works separate prefill and decoding instances 
to improve inference performance and propose a global scheduler to distribute requests. 
Furthermore, Splitwise~\cite{patel2023splitwise} maintains a mixed pool for expanding contracts as needed by the workload 
and adopts a hierarchical two-level scheduling for pool and request management.
TetriInfer~\cite{cunchen2024tetriinfer} utilizes a global scheduler that manages the lifecycle 
of inference requests and a cluster monitor that manages the lifecycle of prefill and decoding instances.
To predict the generation length of LLM inference requests, TetriInfer has fine-tuned a compact LLM model.
MemServe~\cite{hu2024memserve} supports P/D co-located instances apart from prefill and decoding ones, 
with each instance holding a memory pool for memory allocation, index management, and distributed transfer. 
Mooncake~\cite{qin2024mooncake} features a KVCache-centric architecture 
that separates prefill and decoding clusters. 

Although previous works have already 
implemented disaggregated LLMs, 
treating all the prompts in a mixed pool is inadequate 
and the global scheduler easily incurs skewed workload. 
Our work organizes P/D instances in a fine-grained manner (at scale) and adjusts P/D ratio, catering to the changes, 
so that the mismatch between P/D instances is essentially minimized.
Meanwhile, our wok uses an on-demand forwarding upon rejections to balance prefill workload,
and efficient KVCache transfer for faster latency.

\subsection{Serving System at Scale}
Serving LLMs to accelerate the inference and enhance the throughput is widely studied,
such as TensorRT-LLM~\cite{TensorRT-LLM_2023}, 
FastTransformer~\cite{nvidia_fastertransformer_2023}, 
vLLM~\cite{llm_vllm_2023},
DeepSpeed-Inference~\cite{aminabadi2022deepspeed}, 
FlexFlow~\cite{miao2024specInfer}
and TGI~\cite{text_generation_inference_2023}.
The bottlenecks of scaling the system for LLMs are primarily found in two critical areas: 
batch scheduling strategy~\cite{yu2022orca, li2023alpaserve, wu2023fast, aminabadi2022deepspeed, 
hong2024flashdecoding, pope2023efficiently, sheng2024fairness, duan2024muxserve, wu2024loong} and 
memory utilization~\cite{kwon2023pageattention, dettmers2022gpt3, 
dettmers2022spqr, isik2023gptzip, sheng2023flexgen, xiao2023smoothquant}.
Orca~\cite{yu2022orca} introduces iteration-level scheduling to dynamically adjusts the batch size.
FastServe~\cite{wu2023fast} utilizes skip-join Multi-Level Feedback Queue scheduler to minimize job completion time (JCT).
Sheng~\cite{sheng2024fairness} proposes a fair scheduler, the virtual token counter (VTC), 
based on the continuous batching mechanism to achieve fairness in serving.
MuxServe~\cite{duan2024muxserve} proposes a placement algorithm as well as a adaptive batch scheduling strategy 
to serve multiple LLMs concurrently.
PageAttention~\cite{kwon2023pageattention} is widely used for efficient management of xPU memory, 
which dynamically allocates KVCache inspired by the classical virtual memory and paging techniques, 
thereby augmenting the LLM system's throughput.
LLM.int8()~\cite{dettmers2022gpt3}, 
SpQR~\cite{dettmers2022spqr}, 
FlexGen~\cite{sheng2023flexgen} 
and SmoothQuant~\cite{xiao2023smoothquant}
significantly reduce the HBM needed for inference by performing low-bit-precision multiplication.

Benefit from previous technologies, 
the inference in both prefill and decoding (as well as the system throughput) is actually improved.
This paper further studies serving disaggregated LLMs at scale with new challenges.

%% file: tex/extension.tex
\section{Discussion and Extension}
\label{sec:extension}

\subsection{Speculative Decoding}
LLM predicts the next token based on the previous ones in an autoregressive manner.
To accelerate such inference process, 
speculative decoding~\cite{leviathan2023fast, chen2023accelerating} has been proposed. 
This method utilizes a smaller autoregressive model to generate $K$ tokens (educated guesses about future tokens), 
and a larger model decides 
whether to accept all of these tokens or correct them if some are rejected (distribution unchanged~\cite{leviathan2023fast}).

If the autoregressive model is relatively small and the inference latency using CPU is acceptable,
the small model could be entirely managed in the decoding instance of the large model.
However, when the inference latency using CPU is unacceptable,
it has to be treated using NPUs.
Similar to the large model, 
in order to facilitate different batch sizes in P/D and less interruption incurred by P/D mixture,
the small autoregressive model is also disaggregated.
Its prefill part is deployed in the prefill instance of large model,
and its decoding part is deployed in the decoding instance of large model, respectively.
Except for CPU and xPU,
not all types of xPUs can be integrated to support heterogeneous processing for disaggregated LLM directly.
Some are not equipped with RoCE (PCIe is a substitute but with extra overhead).

\subsection{Multi-turn Conversation}
Some works~\cite{gao2024attentionstore, qin2024mooncake} have already considered using host memory for KVCache store.
Our work is orthogonal to such host memory pool.
More specifically, our goal is to improve the hit rate of prefix-aware KVCaches only using HBM
(solved by fine-grained P/D organization per scenario).
With further growth on the number of prefixes and content length (long sequence or multi-turn conversation),
available host memory is useful since its capacity is relatively large.
Although loading KVCache from host (local or remote) incurs extra overhead,
compared with the inference on the entire prompt,
such overhead is gradually acceptable.
Essentially, it is preferred to forward those requests related to the same user or scenario to a subset of prefill instances,
to enhance the hit rate in host memory pool (to the host caches the KV pairs).
Fine-grained P/D organization actually facilitates such demands, 
since it is more likely to hit KVCaches in the same group per scenario
(e.g., pool with multiple tiers to facilitate the same group,
and the gateway can be equipped with the forwarding upon affinity).
Except for the benefits, 
dynamic P/D adjustments (due to scaling, efficiency and recovery) 
further requires the pool to be adjusted for robustness.

%% file: tex/conclusion.tex
\section{Conclusion}
\label{sec:conclusion}
This paper presents the \sysname,
a end-to-end system,
complying with the MLOps,
enables fine-grained P/D organization and adjustment dynamically,
on-demand forwarding with rejections for idle prefill,
and fast block-free KVCache transfer,
to overcome new challenges at scale.
\sysname is implemented and has been deployed 
over tens of thousands of xPUs for more than eight months in commercial use.